\documentclass[acmtog, prologue,dvipsnames, nonacm]{acmart}

\acmSubmissionID{139}

\usepackage{booktabs} 

\usepackage{amsmath}
\usepackage{subfigure}
\usepackage{algpseudocode}
\usepackage{tabularx}
\usepackage{bm}
\usepackage{cancel}
\usepackage{calc}
\usepackage{accents}
\usepackage{tikz}
\usepackage{pifont}
\usepackage{tikz}
\usetikzlibrary{backgrounds}
\usetikzlibrary{calc}
\usetikzlibrary {shapes.geometric} 
\usepackage{wrapfig}
\usetikzlibrary {fit}
\usepackage{cleveref}
\usepackage[shortcuts]{extdash}

\usepackage[ruled]{algorithm2e} 

\SetAlFnt{\small}
\SetAlCapFnt{\small}
\SetAlCapNameFnt{\small}
\SetAlCapHSkip{0pt}

\acmJournal{TOG}

\usepackage{amsmath}
\usepackage[normalem]{ulem}

\newcommand\Revision[1]{{\color{black}{#1}}}

\newcommand{\reffig}[1]{Fig.~\ref{fig:#1}}
\newcommand{\refeq}[1]{Eq.~(\ref{eq:#1})}
\newcommand{\refsec}[1]{Sec.~\ref{sec:#1}}
\newcommand{\reftab}[1]{Table.~\ref{tab:#1}}

\DeclareMathOperator*{\Tr}{Tr}

\DeclareMathOperator*{\argmax}{arg\,max}
\DeclareMathOperator*{\argmin}{arg\,min}

\begin{document}
\title{Shape Space Spectra}

\author{Yue Chang}
\orcid{0000-0002-2587-827X}
\affiliation{%
  \institution{University of Toronto}
  \country{Canada}}
\email{changyue.chang@mail.utoronto.ca}

\author{Otman Benchekroun}
\orcid{0000-0001-6966-5287}
\affiliation{%
  \institution{University of Toronto}
  \country{Canada}}
\email{otman.benchekroun@mail.utoronto.ca}

\author{Maurizio M. Chiaramonte}
\orcid{0000-0002-2529-3159}
\affiliation{%
  \institution{Meta Reality Labs Research}
  \country{USA}}
\email{mchiaram@meta.com}

\author{Peter Yichen Chen}
\orcid{0000-0003-1919-5437}
\affiliation{%
  \institution{MIT CSAIL}
  \country{USA}}
\email{pyc@csail.mit.edu}

\author{Eitan Grinspun}
\orcid{0000-0003-4460-7747}
\affiliation{%
  \institution{ University of Toronto}
  \country{Canada}}
\email{eitan@cs.toronto.edu}

\begin{abstract}


\Revision{Eigenanalysis of differential operators, such as the Laplace operator or elastic energy Hessian, is typically restricted to a single shape and its discretization, limiting reduced order modeling (ROM).}
\Revision{We introduce the first eigenanalysis method for continuously parameterized shape families. Given a parametric shape, our method constructs spatial neural fields that represent eigenfunctions across the entire shape space. It is agnostic to the specific shape representation, requiring only an inside/outside indicator function that depends on shape parameters.}
\Revision{Eigenfunctions are computed by minimizing a variational principle over nested spaces with orthogonality constraints. Since eigenvalues may swap dominance at points of multiplicity, we jointly train multiple eigenfunctions while dynamically reordering them based on their eigenvalues at each step. Through \emph{causal gradient filtering}, this reordering is reflected in backpropagation.}
\Revision{Our method enables applications to operate over shape space, providing a single ROM that encapsulates vibration modes for all shapes, including previously unseen ones. Since our eigenanalysis is differentiable with respect to shape parameters, it facilitates eigenfunction-aware shape optimization. We evaluate our approach on shape optimization for sound synthesis and locomotion, as well as reduced-order modeling for elastodynamic simulation.}

\end{abstract}

%
%


%
%


\begin{teaserfigure}
  \includegraphics[width=\textwidth]{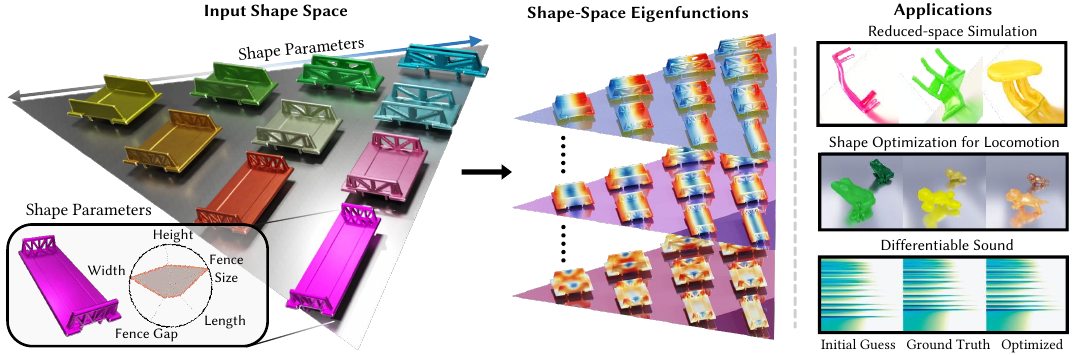}
  \caption{Given a continuously-parameterized shape space \emph{(left)}, we introduce eigenanalysis parameterized over shape space \emph{(center)},  enabling applications such as subspace simulation and inverse design \emph{(right)}. Unlike \Revision{eigenanalysis of a single shape}, our shape-space eigenanalysis is readily differentiable with respect to shape parameters, enabling optimization objectives based on eigenmodes, such as locomotion or timbre. Our focus on shape space in turn requires new techniques for consistency of modes across shape space: compare the shading of shapes within one triangle (eigenfunctions varying over shape space) and across triangles (distinct modes \Revision{revealed} by eigenanalysis).}
  \label{fig:teaser}
\end{teaserfigure}

\maketitle

\section{Introduction}

\begin{figure}[b]
\centering
\includegraphics[width=8cm]{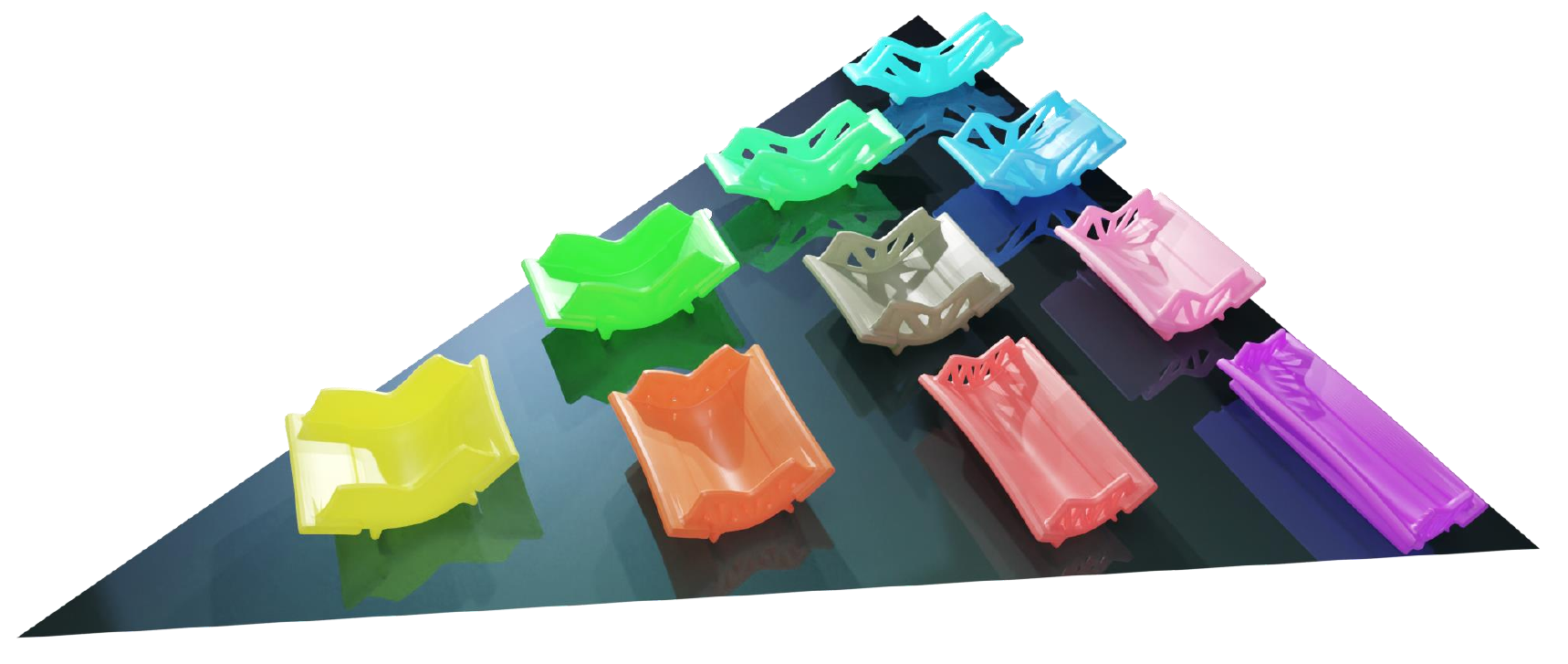}
\caption{
\emph{Elastic eigenfunctions over Shape Space.}
With a single neural model,we can compute eigenfunctions for a bridge over the entire shape-space shown in \reffig{teaser}.
\label{fig:elasticModes}
}
\centering
\end{figure}


\Revision{From the aerodynamics of an airplane wing to the flexibility of a plastic fork, partial differential equations (PDEs) play a crucial role in physics-constrained product design. The eigenfunctions of PDE operators are essential for analyzing design solutions, as they help identify bending or fracture patterns, describe resonant frequencies, and encode geometric properties such as distance and curvature.}


\Revision{However, eigenfunctions are typically computed for a single geometry, and geometry modification requires recomputing the discrete operator and its eigendecomposition, which is nonlinear with respect to geometry. These compute-intensive steps hinder PDE-based shape optimization and interactive design tasks, which often require numerous eigenfunction and derivative evaluations. A key challenge is that evaluating eigenfunctions during optimization is cost-prohibitive, and their derivatives with respect to shape parameters are not readily available.}

\begin{figure}[t]
\centering
\includegraphics[width=8cm]{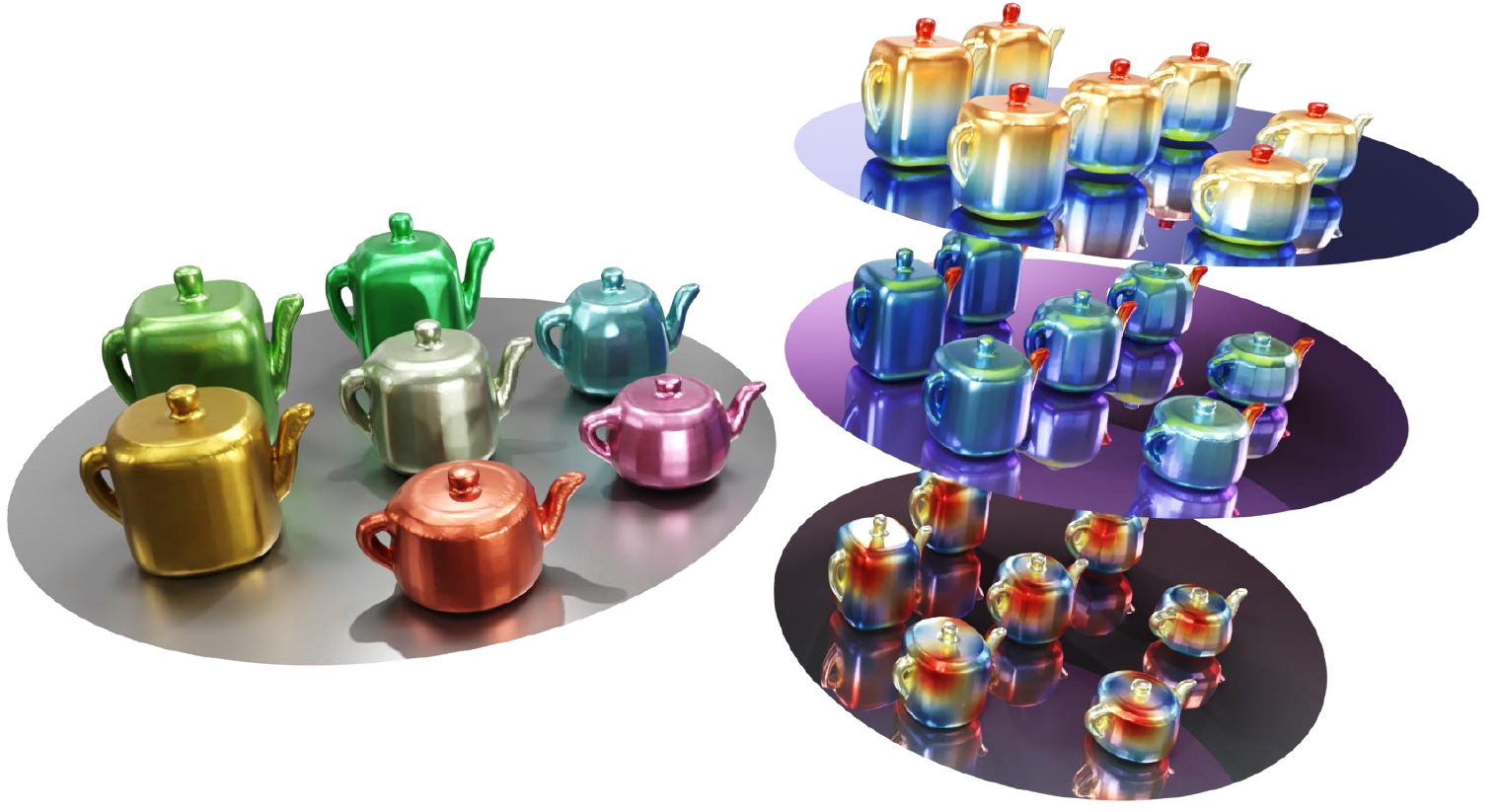}
\caption{
\emph{Laplace Eigenfunctions for a Teapot Shape Space.}
The eigenfunctions of the Laplace operator describe the low-frequency heat distributions. We demonstrate these eigenfunctions across different teapot shapes in the shape space.
\label{fig:teapots}
}
\centering
\end{figure}

\Revision{We introduce a method and representation for eigenmodes over shape space, enabling efficient evaluation for any shape, including those unseen during training (see \reffig{teaser}, \reffig{elasticModes}, \reffig{teapots}). Since our eigenfunctions and eigenvalues share continuous parameters with the corresponding geometry, differentiating eigenfunctions with respect to shape is straightforward. To the best of our knowledge, this is the first proposed method of its kind in the literature.}

\begin{figure}
\centering
\includegraphics[width=8cm]{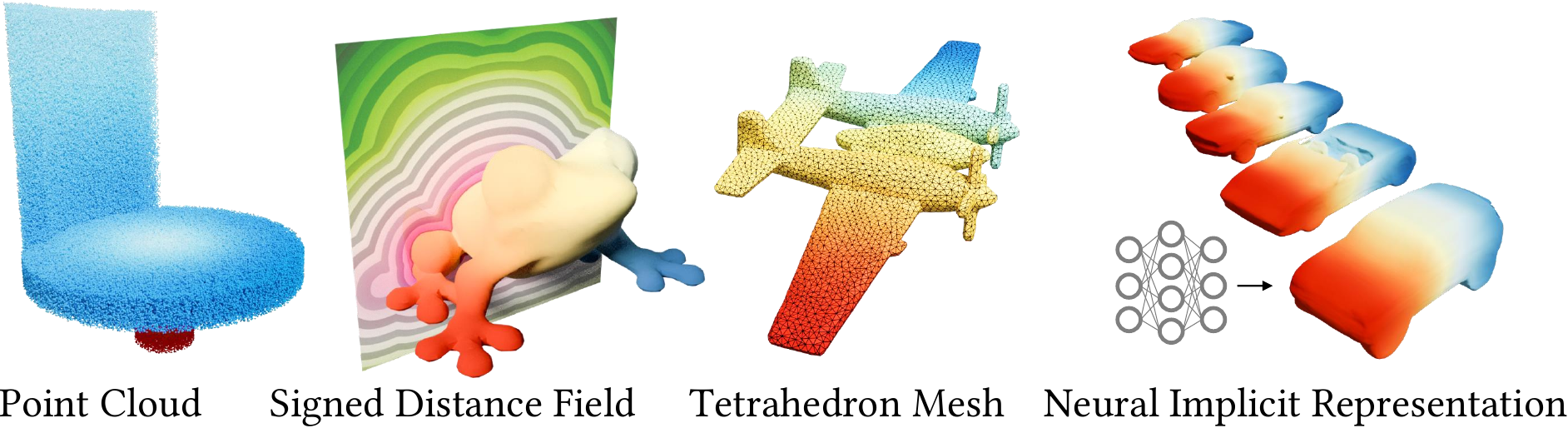}
\caption{
\emph{Eigenfunctions Across Different Representations}
Our method is discretization-agnostic; we have demonstrated the calculation of eigenfunctions for point clouds, signed distance fields, tetrahedral meshes, and neural implicit representations.
\label{fig:discretizationAgnostic}
}
\centering
\end{figure}

%
%
\Revision{Our method is agnostic to the shape space representation (see \reffig{discretizationAgnostic}), requiring only an inside/outside indicator function dependent on shape parameters. As a result, it applies to both manually defined shape interpolation and neural implicit representations trained over real-world datasets \cite{chen2018implicit_decoder}.}

%
%
%
%

\Revision{Our first contribution is a variational method that generates eigenfunctions for single shape domains represented by neural fields. While the eigenfunction problem for PDE operators is well-established in the context of finite element discretizations, it remains largely unexplored for more generalized domains, such as those represented by neural implicit representations, even for single shapes.
To address this, we introduce a projection module at the end of our neural field pipeline, ensuring that each eigenfunction's output lies within the classical eigenfunction constraint space. With these constraints \textit{ensured by construction} in the output space of our network, we can compute the eigenfunctions by minimizing a suitable operator-aware loss functional.}

\Revision{Expanding the aforementioned method to a shape space presents its own unique challenges.}
For a single shape, these eigenfunctions are labeled, computed, and discussed in terms of the order of their dominance, corresponding to a monotonic sequence of eigenvalues.
For a shape space, however, we can no longer think in terms of monotonicity because eigenvalues as functions over shape space often cross and exchange dominance (see Fig.~\ref{fig:compare_sort_new}-right and Fig.~\ref{fig:compare_sort_eigenvalue}-right). 
Maintaining the right ``topological structure'' of eigenfunctions over shape space as they cross each other is critical to engineering design tasks that require differentiating eigenvalues or eigenfunctions with respect to shape parameters. 

Therefore, our second contribution is a method that \emph{jointly} optimizes eigenfunctions over shape space. This requires accounting for the exchange of dominance relationships over shape space and a special filtering step during back-propagation, which we refer to as \emph{causal sorting} and \emph{causal filtering}, respectively.

For the first time, we have a shape-dependent eigenfunction representation that correctly tracks crossovers of eigenvalues at points of multiplicity in shape space.
This approach generalizates across families of shapes, allowing us to predict eigenfunctions for 
new shapes that were not seen during precomputation.
%
%
With its novel dynamic reordering and gradient filtering, it encourages accurate reproduction of eigenfunctions, with consistent modes across different shapes, enabling applications such as single-model multiple-shape subspace physics, warm-starting PDE solutions across shapes, and inverse shape optimization for locomotion and sound profile.

\section{Related Work}
\subsection{Eigenfunctions of PDEs}
Eigenfunctions of PDEs have a broad range of applications such as deformation \cite{skinningcourse:2014,benchekroun2023FastComplemDynamics, James:2002:DyRT, Hildebrandt:2011:ISM}, fluid simulation \cite{Cui:2018:scalableEigenfluid, Witt:2012:eigenfluid}, locomotion \cite{Kry:2009:ModalLocomotion,Nunes:2012:UNVG}, sound analysis \citep{Mark:1966:Drum, Bharaj:2015:metallophone, Brien:2002:RigidSound}, shape analysis \cite{Vallet:2008:SGPMH, Melzi:2018:LMHSSA, Ovsjanikov:2012:FunctionalMaps, Mateus:2008:ASM, Sharma:2010:SMD, Sun:2023:SSCD, Arianna:2019:CFRL}, and geometric deep learning \cite{smirnov2021hodgenet, Sharp:2022:Diffnet}.   

The most straightforward approach to calculating the eigenfunctions for a given shape is to mesh the domain and perform eigendecomposition on the constructed discrete operator. 

For the Laplace-Beltrami operator, despite the availability of various methods \cite{Belkin:2008:DLM, Belkin:2009:PCL, Bobenko2005ADL, Fisher:2006:course, Liu:2014:IDT, Shimada:1995:BubbleMesh, Gueziec:1999:ECNP, Sellan:Overlapping:2019, Sharp:2020:LNT, Pang:2024:NLO}, each with its own advantages, none perfect~\cite{Wardetzky:2007:DLO}, the most commonly adopted discrete operator is the cotangent matrix. 
For elastic energies, the commonly used discrete operator is the elastic Hessian, which varies depending on the specific elastic energy. Commonly used elastic energy models include linear elasticity \cite{Sifakis:2012:FEM}, St. Venant-Kirchhoff \cite{Barbic:2005:stvk}, co-rotational (ARAP) \cite{Rankin:1986:corotational, Sorkine:2007:ARAP}, and Neo-Hookean elasticity \cite{Smith:2018:stableNeohookean}, to name a few. A generalized approach for calculating the Hessian matrix for a given elastic energy can be found in \cite{Kim:2022:tutorial}.

After obtaining the discrete operator, the eigenvectors can trivially be calculated via  eigensolvers  \cite{Duersch:2018:LOBPCG, Arbenz:2005:eigensolve, Nasikun:2022:SubspaceEigen}.
Because all these approaches rely on a discrete operators and perform eigenanalysis on its representative matrix, they are tied to a \emph{single discretization} of \emph{one shape}. 
They cannot find eigenfunctions for general shape \textit{families}.




\subsection{Shape Spaces and Neural Fields}
The shape space in our paper refers to a family of shapes defined by a continuously parameterized shape code. 
Shape spaces can serve various purposes, including defining a solution space for shape optimization \cite{ma2021diffaqua, Jin:2020:diffsound}, visualizing physical properties \cite{Schulz:2017:CAD}, and enabling shape representation and generalization \cite{chen2018implicit_decoder}. 
Common examples of such shape spaces in the graphics community include design spaces \cite{Schulz:2018:AEDT}, shape interpolation \cite{Solomon:2015:wasserstein}, mesh Booleans \cite{Liu:2024:Fuzzy, Yuan:2024:DiffCSG}.
Unfortunately, different shapes usually entail vastly different discretizations, making it very difficult to smoothly navigate geometrically diverse shape spaces.

To make our method discretization-agnostic and able to integrate with large shape spaces, we make use of shape families defined by neural fields \citep{xie2021neural}.
These neural fields parameterize a spatially dependent vector field through a neural network. 
Early seminal efforts by \citet{Park:2019:DeepSDF, chen2019learning, mescheder2019occupancy} utilized this framework for encoding signed distance fields, wherein each distinct latent vector represents a unique geometry. Neural fields have since been extensively applied in various domains, including neural rendering \citep{mildenhall2020nerf}, 3D reconstruction \citep{wang2021neus, yariv2020multiview}, geometry processing \citep{yang2021geometry, aigerman2022neural, Dodik:VBC:2023, mehta2022level, williamson2024neuralgeometryprocessingspherical}, topology optimization \citep{zehnder2021ntopo}, constitutive modeling \citep{li2023neural},  and solving diverse PDE problems \citep{raissi2019physics, chen2022implicit, chen2023crom, chang:2023:licrom, Deng2023}.

Our work builds upon previous research and extends neural fields to model eigenfunctions for PDEs. Neural fields offer a significant advantage over traditional representations like meshes and grids when modeling eigenfunctions: the ability to model a large family of parameterized shapes. Specifically, neural fields enable the modeling of eigenfunctions across a family of shapes by using high-dimensional shape codes as inputs.
%
Furthermore, unlike prior approaches that fit neural fields to pre-existing geometric data, our method trains eigenfunction neural fields in a geometry-informed manner \emph{without} requiring any precomputed eigenfunctions as training data.

\section{Eigenanalysis of a Single Shape}\label{sec:eigen_one_shape}

We begin with defining eigenfunctions for a single shape and then describe our method over shape space in \Revision{ \refsec{eigen_shape_space}}. 
In both cases, we first discuss the variational prespective as a foundation, and then move on to implementation. 
We begin with a variational perspective on eigenanalysis of the Laplace operator for a specific, non-parametric volumetric shape in $\mathbb{R}^n$, before extending to elasticity and shape spaces.

\subsection{Eigenanalysis: A Variational Perspective}

Consider a compact subset \( \Omega \subset \mathbb{R}^n \) with a piecewise smooth boundary
\( \partial \Omega \). We are interested in studying the eigenfunctions of the Laplace 
operator \( \Delta \), subject to appropriate boundary conditions on \( \partial \Omega \).
The Laplace operator is defined as
\[
\Delta u = \nabla \cdot \nabla u = \sum_{i=1}^n \frac{\partial^2 u}{\partial x_i^2},
\]
where \( u \) is a sufficiently smooth function defined on \( \Omega \).

The dominant eigenfunction $\phi_1(\bm{x})$ minimizes Dirichlet energy
\begin{align} \label{eq:dirichlet}
E_D[\phi] = \frac{1}{2} \int_{\Omega} |\nabla \phi|^2 \, d\Omega,
\end{align}
among $\mathcal{U} = \{ f \in L^2(\Omega) \mid \|f\|_2 = 1 \}$, the unit-norm square-integrable functions in $\Omega$. Restricting the search to unit-norm functions (akin to taking the Rayleigh quotient) helps canonize the minimizer and equates eigenvalue to Dirichlet energy, $\lambda_1 = E_D[\phi_1]$.

\paragraph{Peeling off eigenfunctions}
The subdominant eigenfunction \emph{also} minimizes Dirichlet energy, but this time in the space $\phi_1^{\perp}$, the \emph{orthogonal complement} to $\phi_1$ in $S$. And so forth, in order of dominance: the $i$'th eigenfunction $\phi_i$ minimizes Dirichlet energy $\lambda_i = E_D[\phi_i]$ in $\mathcal{C}_i=\text{span}\{\phi_1, \ldots, \phi_{i-1}\}^\perp$, the space orthogonal to earlier modes:
\begin{align} \label{eq:variational-peeling}
\phi_i = \argmin_{\phi \,\in\, \mathcal{U} \,\cap\, \mathcal{C}_i} E_D[\phi] \ .
\end{align}
In this iterative ``peeling'' procedure, echoing the Courant\-/Fischer\-/Weyl min\-/max principle, each dominant eigenfunction is found and ``peeled away'' revealing the complementary subspace containing the subdominant eigenspace, where the procedure is repeated. Because the spaces are nested, the minimizer of the ``bigger'' space cannot be greater than subsequent ``smaller'' spaces, therefore $\lambda_1 \leq \lambda_2 \leq \ldots\ $ .

\begin{figure}
\centering
\includegraphics[width=8cm]{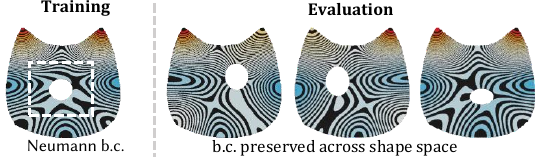}
\caption{
\emph{Boundary Condition.} Minimizing the Dirichlet energy on a shape naturally enforces a vanishing Neumann boundary condition. To illustrate this, we visualize the isolines of the eigenfunctions. Notably, the gradients along the boundary are nearly zero, highlighting the fulfillment of the Neumann condition.
} 
\label{fig:boundary_condition} 
\centering
\end{figure}

Since we have not explicitly enforced any boundary conditions, the minimizers all satisfy the \emph{natural} condition, which, for the Dirichlet energy, is the vanishing Neumann value $\frac{\partial \phi_i}{\partial n} = 0$ on $\ \partial \Omega$, as shown in \reffig{boundary_condition}. \Revision{We discuss avenues for essential boundary conditions in \refsec{discussion}.}

\begin{figure}
\centering
\includegraphics[width=8cm]{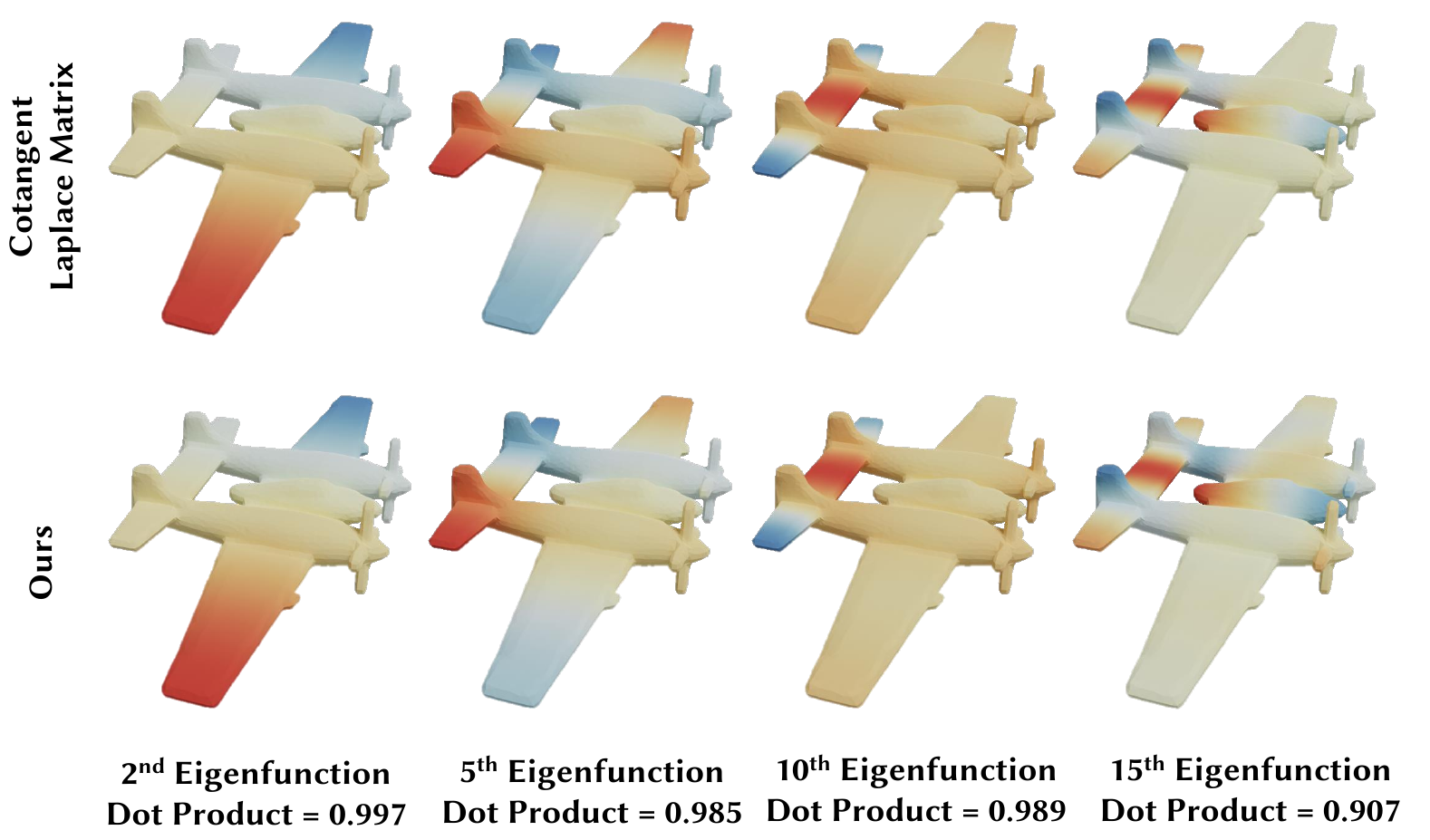}
\caption{
\emph{Comparison with eigenfunctions from cotangent Laplacians.}
When training on a single shape, our method converges to results consistent with traditional eigenanalysis. We compare the eigenfunctions obtained using our approach with those derived from a cotan Laplace (linear finite element on triangle mesh) matrix. The accuracy of our method matches that of existing techniques that rely on matrix construction from point cloud sampling.
\label{fig:showErrorShape}
}
\centering
\end{figure}

\subsection{Implementation with Neural Fields}\label{subsec:oneshapeimplementation}

We implemented the optimization described in Equation \ref{eq:variational-peeling} using neural fields. Our results show good agreement with prior discrete Laplace operators (see \reffig{showErrorShape}), boundary condition satisfaction (see \reffig{boundary_condition}), and SE(3) (or SE(2)) invariance (see \reffig{se3}).

Our implementation models each eigenfunction $\phi_i$ as the composition of a corresponding neural field and a projection operator: 
\begin{align} \label{eq:inference-then-project}
\phi_i = \mathcal{P}_i \circ \overline{\phi}_i \ .
\end{align}
The neural field $\overline{\phi}_i$ is a multilayer perceptron (MLP) mapping domain position, $\bm{x} \in \Omega$, to field value, $\overline{\phi}_i(\bm{x})$. The projection operator $\mathcal{P}_i : L_2 \rightarrow \mathcal{U} \cap \mathcal{C}_i$ maps any field to a constraint-satisfying field.

Algorithm \ref{alg:single_shape} performs eigenanalysis for the Laplace operator over a single shape. We describe the optimization, energy (loss) evaluation, and projection operator in turn.

\begin{algorithm}[t]

$epoch$ = 0\; 

\Repeat{$epoch = MaxEpoch$}{
        $\mathcal{X}$ = $\{\bm{x}_1,\ldots\}$; \Comment{Sample Domain $\Omega$} \;
        $\phi_{prev}(\mathcal{X})$ = Ones like $\mathcal{X}$ \; 
        \Comment{Hardcode the known eigenfunction}\; 
        \For{$i$ in range $[0,k)$}
        {
      Evaluate Network $\overline{\phi}_{i}(\mathcal{X})$\;
      Calculate {$\bm{\lambda}$} by doing the projection in Equation \ref{eq:projections}\;
      Calculate $\phi^p_{m+1}(\mathcal{X})$ by Equation \ref{eq:project_out}\;
      Calculate $\phi(\mathcal{X})$ by Equation \ref{eq:normalization}\;
      \For{$\bm{x}$ in $\mathcal{X}$ = $\{\bm{x}_1,\ldots\}$}
      {
      Evaluate gradient $\frac{\partial \phi(\bm{x})}{\partial \bm{x}}$ by Equation \ref{eq:normalization_gradient}\;
      \Comment{In practice, this is implemented using tensors}\; 
      }
      Calculate loss $\mathcal{L}$ by Equation \ref{eq:loss} \;
      Backward loss \;
      Concatenate $\phi_{prev}(\mathcal{X})$ with $\phi_{i}(\mathcal{X})$ and detach\; 
        \Comment{Update for future orthogonal constraints}\;  
     }
     $epoch$ = $epoch+1$ \; 
      }

\caption{Optimization on a Single Shape for Laplace Eigenfunction}
\label{alg:single_shape}
\end{algorithm}


Each neural field, $\overline{\phi}_i$, is trained by minimizing the loss $\mathcal{L} = E_D[\phi_i]$. After convergence, the next neural field is trained, in sequence. We estimate the domain integral with stochastic cubature:
\begin{align} \label{eq:loss}
\tilde{\mathcal{L}} =  \tilde{E}_D[\phi_i] = \sum_{\bm{x_j}\in \mathcal{X}} |\nabla \phi_i(\bm{x_j})|^2 \ .
\end{align}
Our implementation uniformly samples cubature points $\mathcal{X} = \{x_j \in \Omega\}$ via rejection sampling, drawing from a uniform distribution over an axis-aligned bounding volume, and rejecting samples based on the indicator (inside/outside query) function of $\Omega$. Since our implementation assumes only an indicator function, it is agnostic to the representation of $\Omega$.

\label{sec:transformations}


The projection $\mathcal{P}_i$ takes a general field $\overline{\phi}_i$ and returns
the closest field $\phi_i \in \mathcal{U} \,\cap\, \mathcal{C}_i$ satisfying the orthogonality and unit-norm conditions. The projection is composed of two steps: orthogonalization and normalization, which we represent via the dataflow schematic:
\begin{align*}
     \mathcal{P}_i: \quad \overline{\phi}_i \xrightarrow{\textrm{Gram–Schmidt}} \phi^p_i \xrightarrow{\textrm{normalize}} \phi_i \ .
\end{align*}

\begin{figure*}
\centering
\includegraphics[width=\textwidth]{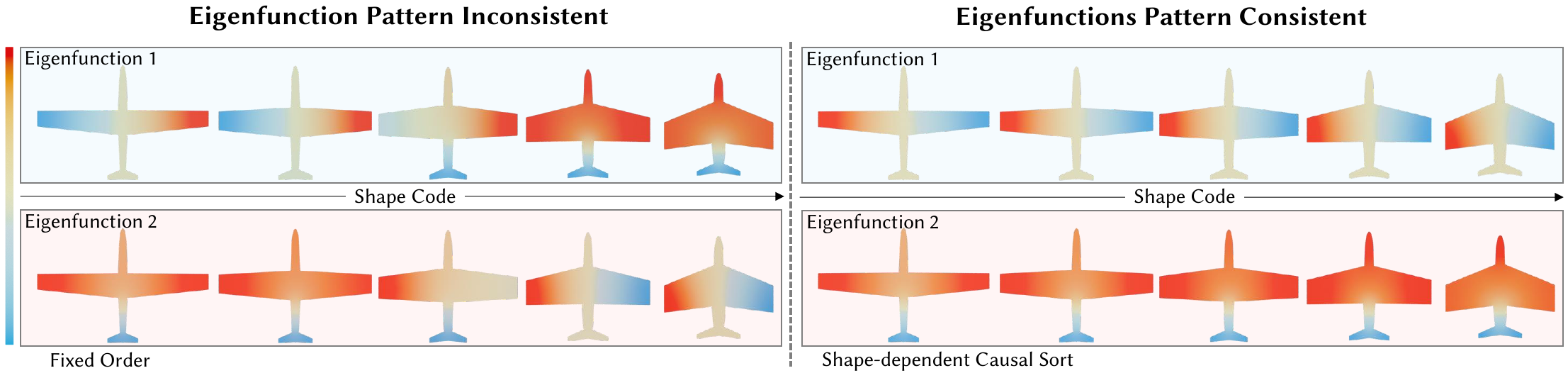}
\caption{
The wing shape varies over shape space. \emph{(left)} \textbf{Ablation:} eigenfunctions 1 \& 2 swap modal patterns at the point of eigenvalue multiplicity (see \reffig{compare_sort_eigenvalue}). \emph{(right)} \textbf{Our approach:} shape-dependent causal sorting improves mode consistency across shape space.
\label{fig:compare_sort_new} 
}
\centering
\end{figure*}

\paragraph{Orthogonalization (projection to $\mathcal{C}_i$)} 


Any two eigenfunctions $\phi_i$ and $\phi_j$, $i \neq j$, must be orthogonal, $\int_{\Omega}\phi_i \phi_j d\Omega = 0$, a constraint we achieve by Gram–Schmidt othogonalization. Given a candidate (unconstrained) function $\overline{\phi}_m$ not yet orthogonal to all previous eigenfunctions, we find its projection ${\phi}^p_m$ onto the orthogonal subspace by subtracting from $\overline{\phi}_m$ the component already spanned by the $m-1$ dominating eigenfunctions $\bm{\phi} = (\phi_1, \dots, \phi_{m-1})$. We seek the $\bm{\lambda}$-weighted linear combination of dominating eigenfunctions $\bm{\phi}$ that best approximate $\overline{\phi}_m$,
\begin{align} \label{eq:projections}
\bm{\lambda} = \argmin_{\bm{\lambda} \in \mathbb{R}^{m-1}} 
\left\|\bm{\lambda}^T\bm{\phi} - \overline{\phi}_m\right\|_2 ,
\end{align}
where $\|\cdot\|_2$ is the $L_2$ norm, which, like all domain integrals, we estimate by uniform stochastic cubature.
Next, we remove the component:
\begin{align} \label{eq:project_out}
\phi^p_m = {\overline{\phi}}_m - {\bm{\lambda}^T\bm\phi} \ .
\end{align}
The gradient, required for training, follows by chain rule.

\paragraph{Normalization (projection to $\mathcal{U})$} 

We enforce the unit-norm constraint by normalizing ${\phi}^p(\bm{x})$:
\begin{align} \label{eq:normalization}
\phi(\bm{x}) = \frac{{\phi}^p(\bm{x})}{\|\phi^p\|_2} \ ,
\end{align}
again \Revision{estimating} the $L_2$ norm by uniform stochastic cubature.

\Revision{To accelerate gradient computation, we approximate the gradient of the norm by treating it as independent of its argument, a similar approach to \citep{williamson2024neuralgeometryprocessingspherical, Levy:2010:SMP}.
}
\begin{align} \label{eq:normalization_gradient}
(\partial/\partial \bm{x}) \phi \approx \|\phi^p\|^{-1}_2  
(\partial/\partial \bm{x}) {\phi}^p\ .
\end{align}

Now, we have successfully constructed a set of continuous functions that are orthogonal to each other and possess a unit norm.

\paragraph{Implementation details for a single Shape}
We use ADAM \cite{Kingma:2017:adam} to optimize the neural network weights, and implement our method in PyTorch, evaluating spatial gradients such as $\nabla \overline{\phi}_i(\bm{x})$ using PyTorch's autodiff. Theory suggests that to minimize bias in the optimization, each domain integral should be estimated using an independently sampled cubature set, however, we found that using one cubature set per epoch, for all integrals, provides good results.

\paragraph{Extension to Elasticity}
\label{sec:elas}\label{sec:elasticity}
Our method can also be applied to elasticity. In this case, the resulting eigenfunctions (elastic modes) are vector-valued fields, $\boldsymbol{\phi}_i(\bm{x}) \mapsto \mathbb{R}^3$. We implemented linear elastic energy \cite{Sifakis:2012:FEM}. The eigenfunctions in this context are minimizers of the energy functional:
\begin{align} \label{eq:elas}
E_e[\boldsymbol{\phi}] = \frac{1}{2} \int_{\Omega} \mu |\nabla \boldsymbol{\phi} + \nabla \boldsymbol{\phi} ^ T|_{F}^2 +\frac{\lambda}{2} {\Tr} ^2 (\nabla \boldsymbol{\phi} + \nabla \boldsymbol{\phi} ^ T) d\Omega,
\end{align}
Here, $\boldsymbol{\phi}$ represents the vector-valued eigenfunction, and $\nabla \boldsymbol{\phi}$ is the deformation gradient (or Jacobian), expressed as a $3 \times 3$ matrix. The term $\nabla \boldsymbol{\phi} + \nabla \boldsymbol{\phi}^T$ is known as the small strain tensor. The Frobenius norm is denoted by $|\cdot|_F$, and $\mu$ and $\lambda$ are the Lamé coefficients. We estimate $\tilde{E}_e \approx E_e$ using uniform stochastic cubature.

Crucially, other than substituting $\tilde{E}_D$ with $\tilde{E}_e$, and reconsidering the known modes (discussed below), our theory and implementation are unchanged, highlighting the broader applicability of the method.

\paragraph{Handling Known Modes}
When subject to Neumann boundary conditions, the dominant mode ($\lambda_1 = 0$) of the Laplace operator is always the constant function, irrespective of domain geometry. 
While the peeling implementation described above is able to find this first trivial mode, this is wasted computation, since the result is already known. 
Therefore, we hard-code the known geometry-independent eigenfunction analytically, and represent the remaining, geometry-dependent modes using neural fields. 
For the elasticity operator, we hard-code the zero eigenvalue modes corresponding to rigid translations and rotations.

\section{Eigenanalysis over Shape Space}

\label{sec:eigen_shape_space}
We are ready to dive into training over shape spaces. Let the 
domain $\{\Omega^{\bm{g}}|\bm{g}\in\mathcal{D} \}$ be 
parameterized by a geometry code $\bm{g} \in \mathcal{D}$ drawn from a shape space  $\mathcal{D}$. Since choosing a point $\bm{g} \in \mathcal{D}$ fixes the shape of the domain $\Omega^{\bm{g}}$, we could use the \Revision{eigenanalysis of a single shape} from \S\ref{sec:eigen_one_shape} to determine the eigenfunctions $\phi^{\bm{g}}_i(\bm{x})$.

On the other hand, by explicitly leaving $\bm{g} \in \mathcal{D}$ as a free parameter, we can think of eigenfunctions $\phi^{\bm{g}}_i(\bm{x}) \equiv \phi_i(\bm{g}, \bm{x})$ as spatial fields parameterized over shape space, whose cross-sections at some $\bm{g} = \bm{g}_j$ correspond to the eigenfunctions over $\Omega^{\bm{g}_j}$.  Classical results from perturbation theory~\cite{Kato:1980:PerturbationTF} state that for small, smooth perturbations of the domain, eigenvalues and eigenfunctions of elliptic operators (e.g., the Laplacian) vary smoothly, or even analytically. This opens the door to a reduced-order or parametrized representation of the eigenfunctions across the entire shape space $\mathcal{D}$, that is, to efficient learning and representation of eigenfunctions $\phi^{\bm{g}}_i(\bm{x}$) across a continuous family of domain geometries.

\subsection{Shape-Space Eigenanalysis: A Variational Perspective} \label{sec:from-sequence-to-set}

A first, albeit misguided, attempt to generalize the method discussed in \S\ref{sec:eigen_one_shape} might be to modify the domain of integration to be the product of shape space and the spatial domain. The first eigenfunction $\phi_1$ minimizes the integral of Dirichlet energy over shape space $\int_{\mathcal{D}} E_D[\phi^{\bm{g}}_1]  d\mathcal{D}$, the second eigenfunction $\phi_2$ does so restricted to $\phi_1^{\perp}$, and so forth. But such peeling would inherently yield eigenfunctions that maintain a fixed dominance relationship across shape space, which, as we are about to see, is damaging and unnecessary.

We present a didactic example in \reffig{compare_sort_new}-left, where the shape space represents a family of airplanes of changing wing thicknesses $\bm{g}$. As the shape code changes, the vertically oriented eigenfunctions (varying from the left wingtip to the right wingtip) and the horizontally oriented eigenfunctions (varying from the tail to the aircraft's body) swap dominance. If we also plot the eigenvalues of the two eigenfunctions, as shown in \reffig{compare_sort_eigenvalue}-left, we observe that the crossover occurs at a point of geometric symmetry and eigenvalue multiplicity.

In this example, the classical view of eigenfunctions ordered by dominance leads to defining the blue curve that (by construction) always dominates the red curve throughout shape space \reffig{compare_sort_eigenvalue}-\emph{left} (``Fixed Order''). Observe how the \Revision{blue} curve represents an eigenfunction that is only \emph{piecewise} smooth over design space, with a kink at the crossover; the same for the red curve. Such a kink is undesirable for neural field training and its applications, consuming more network capacity, slowing convergence, and leading to numerical challenges or failures in applications that harness the smoothness and differentiability of eigenfunctions with respect to $\bm{g}$. Worse, such discontinuities increase in number and topological complexity with the dimension of design space and the number of eigenfunctions. 

Fortunately, these discontinuities are an unnecessary, fictitious fabrication arising only from clinging to the single shape mindset. Indeed, \citet{Kato:1980:PerturbationTF} argues instead for viewing eigenvalues as an \emph{unordered set}, whose subscript indexing merely provides a unique identifier, not an ordering. Now eigenvalues and eigenfunctions are smooth even over crossings at points of multiplicity. This is depicted in \reffig{compare_sort_new}-\emph{right} and \reffig{compare_sort_eigenvalue}-\emph{right} (``Shape-Dependent Causal Sort''), wherein the eigenvalue curve colored blue corresponds to the eigenfunction that is smooth and horizontally-oriented \emph{throughout} shape space; likewise, the red curve corresponds to the vertically-oriented mode smoothly-varying over shape space. Fig.~\ref{fig:chair_crossover} depicts the same kind of comparison for a complex shape space.

We therefore turn to a variational principle that \emph{jointly} considers a \emph{set} of eigenfunctions minimizing the sum of their Dirichlet energies integrated over shape space,
\begin{align}  \label{eq:shape-space-variational-principle}
\argmin_{\phi_0\ldots \phi_k} \sum_{i=0}^k \int_{\mathcal{D}} E_D[\phi^{\bm{g}}_i] \, \mathrm{d}\bm{g} \ , \quad \textrm{subject to ``orthogonality.''}
\end{align}
This energy is a straightforward extension of the single shape case, and the story would end here if eigenvalues never crossed. The key remaining ingredient is the enforcement of orthogonality relationships, which becomes nontrivial because dominance relationships vary over shape space. 

\subsection{Implementation Using Neural Fields}

\begin{figure}
\centering
\includegraphics[width=8cm]{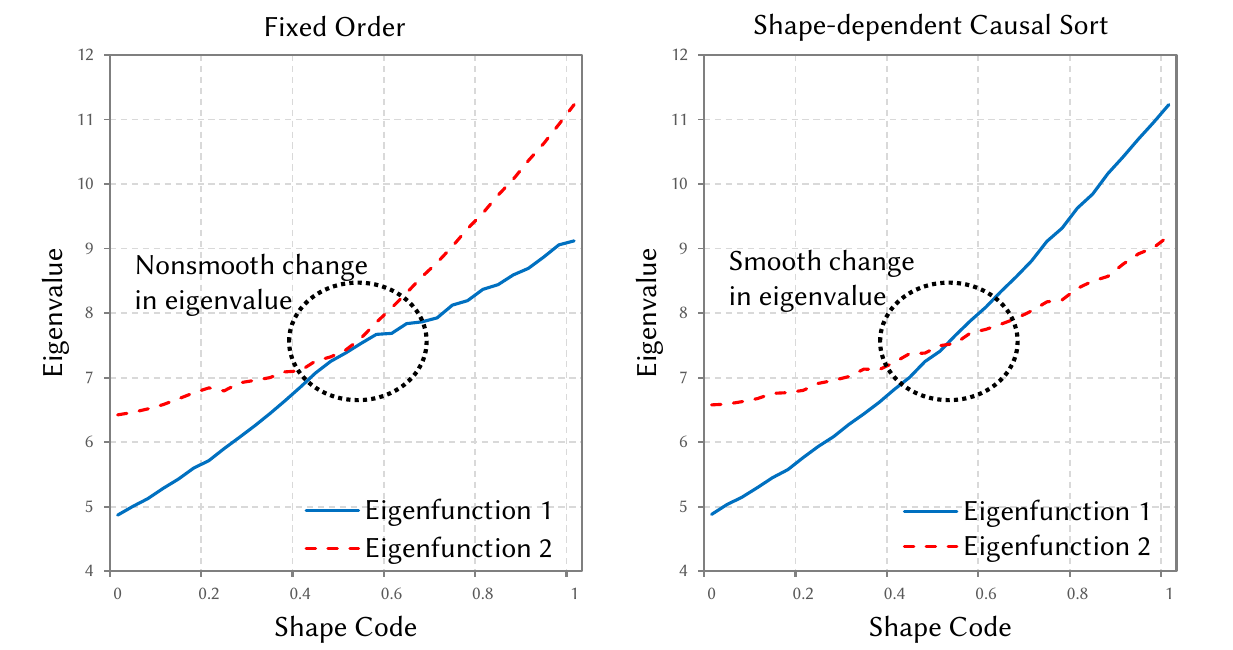}
\caption{\Revision{We visualize the eigenvalues corresponding to ~\reffig{compare_sort_new}. The x-axis shape code matches the middle panel of ~\reffig{compare_sort_new}, following the same shape variation.} The (in)consistency of modal patterns in~\reffig{compare_sort_new} can be understood by examining the eigenvalues as functions over shape space. \emph{(left)} \textbf{Ablation:} Comparing the eigenvalue plot to the eigenfunctions depicted in ~\reffig{compare_sort_new}-\emph{left}, the mode swap occurs at the point of multiplicity, where the eigenvalue curve kinks. \emph{(right)} \textbf{Ours:} Our  eigenmodes are as consistent as possible across shape space, and our eigenvalue  curves are smoother, crossing as appropriate at points of multiplicity.
\label{fig:compare_sort_eigenvalue} 
}
\centering
\end{figure}

To achieve the desirable construction of \reffig{compare_sort_eigenvalue}-\emph{right}, we must forgo sequential peeling of an ordered sequence and instead jointly optimize an unordered set of eigenfunctions. As depicted in the same figure, our optimization must allow different functions to dominate in different regions of shape space. 

This novel direction presents three interwoven technical challenges: (1) We seek to find multiple eigenfunctions governed by coupled energy minimization principles across a descending chain of spaces, (2) Each minimization problem is constrained to a subspace determined by the solution of all preceding minimizations, establishing a causal relation, and (3) The ordering of these causal relations varies across the shape space.



To address these challenges, we extend the single shape algorithm with three novel, interconnected concepts, all three \emph{strictly required} to obtain eigenfunctions analytic over shape space:
\begin{enumerate}
    \item \textbf{Joint training}: As we have seen, sequential peeling produces an incorrect structure because no single ordering of eigenfunctions is valid across all of the shape space. Therefore, we \emph{must} learn $n$ eigenfunctions \emph{jointly}.
    \item \textbf{Gradient causal filtering}: As we shall see, a na\"{\i}ve approach to joint learning suffers from an action-reaction artifact, whereby functions earlier in the causal chain are affected by an orthogonality constraint that should only affect functions subsequent in the chain. To address this, 
    our backpropagation filters the gradient to enforce the causality of the orthogonality constraint.
    \item \textbf{Shape-dependent causal sorting}: Since the ordering of causal relations cannot be predetermined and indeed varies over shape space, we determine the order dynamically. At each evaluation of the loss function, we re-establish the causal ordering of orthogonality constraints based on the relative dominance of eigenvalues.
\end{enumerate}
\textbf{Necessity of these three advances:} These three interwoven concepts are all \emph{required} for our method to achieve, for the first time, a shape-dependent eigenfunction representation that correctly tracks crossovers of eigenvalues at points of multiplicity.

 We show that our dynamic reordering encourages smoother eigenfunctions across different shape (see \reffig{compare_sort_new}-right) and facilitating eigenfunction-dependent shape design. 


As in \S\ref{subsec:oneshapeimplementation}, we represent each eigenfunction $\phi_i$ as the composition of a corresponding neural field and a projection operator, $\phi_i = \mathcal{P}_i \circ \overline{\phi}_i$. However, 
this time the neural field maps \emph{both} shape code, $\bm{g} \in \mathcal{D}$, and domain position, $\bm{x} \in \Omega$, to field value, $\overline{\phi}_i(\bm{g},\bm{x})$; and the projection operator, $\mathcal{P}_i$, which restrict the field to the constraint-satisfying function space $\mathcal{U} \cap \mathcal{C}_i$, must now account for the variation of dominance relations over shape space.

Algorithm \ref{alg:multiple_shape} performs eigenanalysis for the Laplace operator over shape space. We describe the optimization, energy (loss) evaluation, and projection operator in turn.

\subsubsection{Energy Evaluation}\label{sec:imp}


A set of neural fields $\{\overline{\phi}_0, \ldots, \overline{\phi}_k\}$ is jointly trained by minimizing the loss (recalling \refeq{shape-space-variational-principle})
\begin{align} 
\mathcal{L} = \sum_{i=0}^k \int_{\mathcal{D}} E_D[\phi^{\bm{g}}_i] \, \mathrm{d}\bm{g} \ . 
\end{align}
We estimate integrals over shape space and spatial domains via uniform stochastic cubature (see \S\ref{subsec:oneshapeimplementation}), uniformly drawing a finite set of shapes $\mathcal{G}=\{\bm{g}_j \in \mathcal{D}\}$, yielding the discretized loss
\begin{align} 
\tilde{\mathcal{L}} = \sum_{i=0}^k \sum_{\bm{g}_j \in \mathcal{G}} \tilde{E}_D[\phi^{\bm{g}_j}_i] \ .
\end{align}
The computation of the loss includes evaluation of $\tilde{E}_D$, wherein the gradient $\nabla \equiv \nabla_{\bm{x}}$ is a spatial gradient calculated solely with respect to spatial coordinates (recall \refeq{loss}). The implementation for elasticity simply replaces $\tilde{E}_D$ with $\tilde{E}_e$ (see \S\ref{sec:elasticity}).


\subsubsection{Joint Training and Gradient Causal Filtering}  \label{sec:causality}

Joint training of eigenfunctions complicates the enforcement of orthogonality constraints. Since the weights of both the dominating and dominated neural fields are now optimization variables, backpropagation produces an undesirable artifact. The gradient of the orthogonalization step has components along both the dominating and dominated fields. This is the ``action-reaction'' principle of a constraint force. A na{\"\i}ve backpropagation of this constraint gradient disrespects the direction of causality, with the dominated field ``pushing back'' the dominating field away from its optimum. 
\begin{wrapfigure}{l}{0.2\textwidth}
    \centering
    \includegraphics[width=0.24\textwidth]{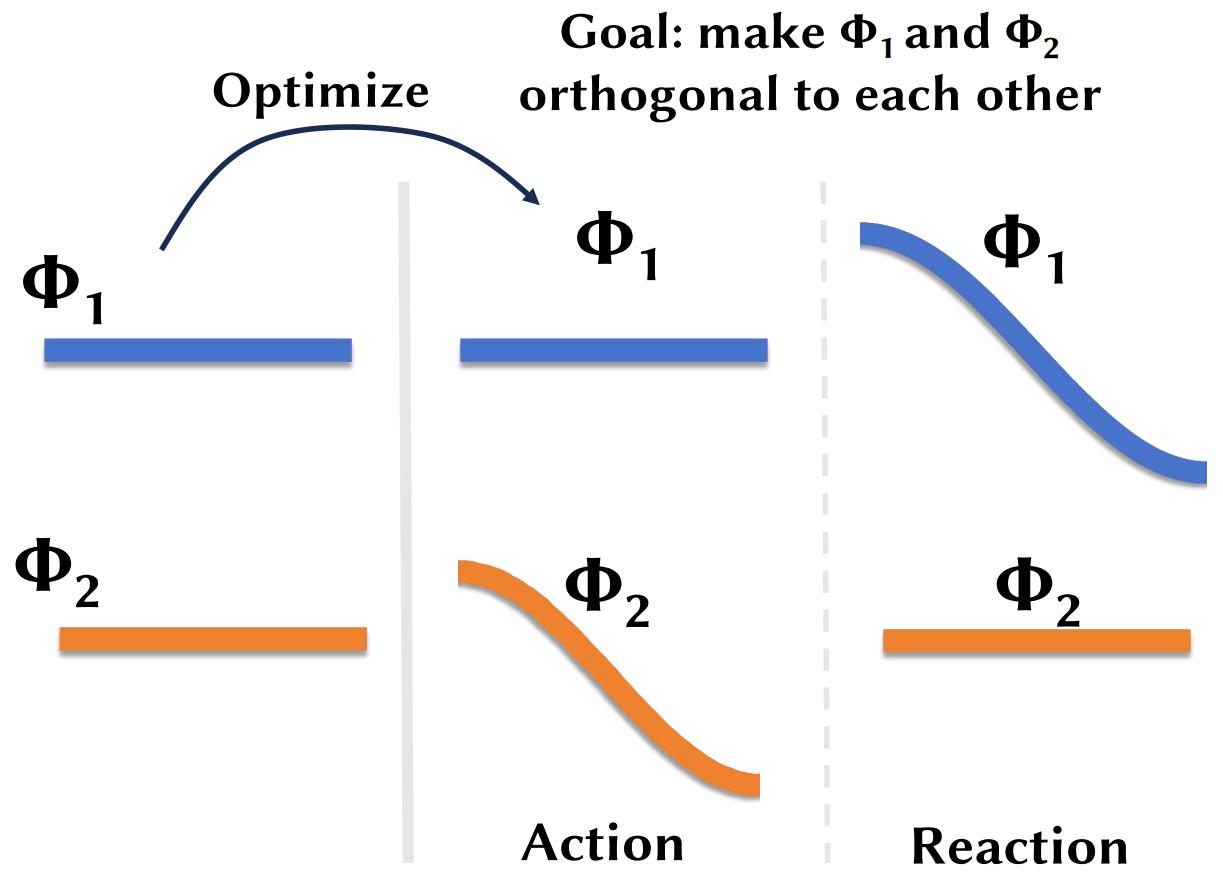}
\end{wrapfigure}
The \Revision{inset} figure illustrates the action-reaction artifact. Given a dominating function $\phi_1$ and a dominated function $\phi_2$ subject to the constraint $\phi_2 \in \phi_1^{\perp}$, there are two ways to decrease the Dirichlet energy of the dominated function, either (1) by making adjustments within the orthogonal subspace $\phi_1^{\perp}$, or (2) by making adjustments to $\phi_1$ so as to modify the admissible space $\phi_1^{\perp}$. The first option (``action'') respects causality, whereas the latter (``reaction'') does not; yet both arise from differentiating the constraint $\phi_2 \in \phi_1^{\perp}$ with respect to $\phi_1$ and $\phi_2$.

\begin{figure}
    \centering
    %
    %
    \includegraphics[width=\linewidth]{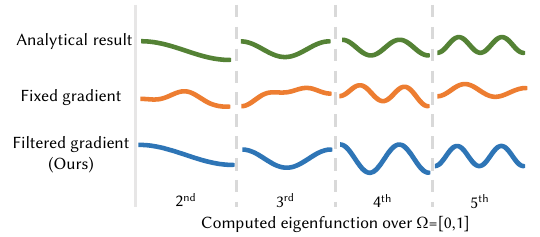}
    \caption{To illustrate the importance of \emph{gradient causal filtering}, we  compare Laplacian eigenfunctions over the unit interval as produced by: \emph{(top)} \textbf{Ground truth:} analytical evaluation of sinusoids satisfying the vanishing Neumann condition; \emph{(middle)} \textbf{Ours:} variational eigenanalysis including gradient causal filtering; 
    \emph{(bottom)} \textbf{Ablation:} our approach excluding gradient causal filtering.}
    \label{fig:causal-sort}
\end{figure}

To eliminate the causality-violating reaction, we must \emph{not} differentiate the constraint with respect to the dominating function. We call this causality-enforcing ignoring of a gradient term \textbf{gradient causal filtering}. Such filtering could be used to enforce any causal constraint relationship in joint training of neural networks, and we use it to protect the causality of the descending chains of orthogonal spaces. We implement the filtering using \colorbox{gray!20}{\texttt{detach()}} in PyTorch. After detachment, the gradient of the orthogonality constraint is considered \emph{only} with respect to the dominated eigenfunction. Figure~\ref{fig:causal-sort} presents an ablation study for gradient causal filtering.

\paragraph{Speedup of single shape training.}
We train jointly because it is the only viable path to producing correct results over shape space, but as a side bonus, the joint approach trains $2\times$ to $4\times$ faster, too, compared to a sequential approach.

\vspace{-0.2cm}
\subsubsection{Shape-Dependent Causal Sorting}  \label{sec:sort}

Since fixing the dominance order of the eigenfunctions is undesirable (recall \S\ref{sec:from-sequence-to-set}), our optimization allows ordering to
vary over shape space.
To achieve the desirable results of Fig.~\ref{fig:compare_sort_new}-\emph{right}, we determine the dominance order \emph{dynamically} at each optimization step, by comparing the eigenvalues (equivalently, Dirichlet energy) of the two eigenmodes for some point in shape space. Wherever the horizontally-oriented eigenfunction has the smaller eigenvalue, it dominates the vertically-oriented eigenfunction, and vice-versa.

Therefore, the order of projection in $\mathcal{P}$ is determined by the eigenvalue and varies for each shape. Since we integrate over shape space using stochastic cubature, this amounts to determining the dominance chain at each sample point in shape space by sorting the eigenvalues (i.e., Dirichlet energy) of the eigenfunctions. Algorithm~\ref{alg:multiple_shape} first evaluates the energy for each unit-norm eigenfunction, then sorts to determine the causal ordering, and then applies this \Revision{order to construct} the projection operator.

With dynamic sorting, we  produce the desirable results of Fig. ~\ref{fig:compare_sort_new}-right and ~\ref{fig:compare_sort_eigenvalue}-right, where each eigenfunction and eigenvalue evolves smoothly over shape space.

\begin{algorithm}[t]

$epoch$ = 0\; 

\Repeat{$epoch = MaxEpoch$}{
        $\bm{g}$ = Sample Geometry Code\;
        \Comment{The geometry code is an explicit shape parameter (e.g., width) or a latent variable (e.g., from an auto-decoder).}\;
        $\bm{x}$ = Sample Domain $\Omega$\;
        $E_{d}$ = []\;
        \For{$i$ in range $[0,k)$
    }{
      Evaluate Network $\overline{\phi}_{i}(\bm{x}, \bm{g})$ and gradient $\frac{\partial\overline{\phi}_{i}(\bm{x}, \bm{g})}{\partial \bm{x}}$\;
      $E_{d}$.append($\frac{1}{\sum\overline{\phi}_{i}(\bm{x}, \bm{g})^2}\sum(\frac{\partial\overline{\phi}_{i}(\bm{x}, \bm{g})}{\partial \bm{x}})^2$) \;
      \Comment{Evaluate Dirichlet energy on normalized gradients}\; 
        
     }
     index, sortedEigenvalue = sort($E_{d}$)

     \For{$idx$ in range $[0,k)$
    }{
    $i$ = index[$idx$]
    \Comment{Use the order from the sorted index}\;

    Do projection and optimization same as Algorithm \ref{alg:single_shape}\;

    }
     $epoch$ = $epoch+1$ \; 
      }

\caption{Training Over a Shape Space}
\label{alg:multiple_shape}
\end{algorithm}

\begin{figure}
\centering
\includegraphics[width=8cm]{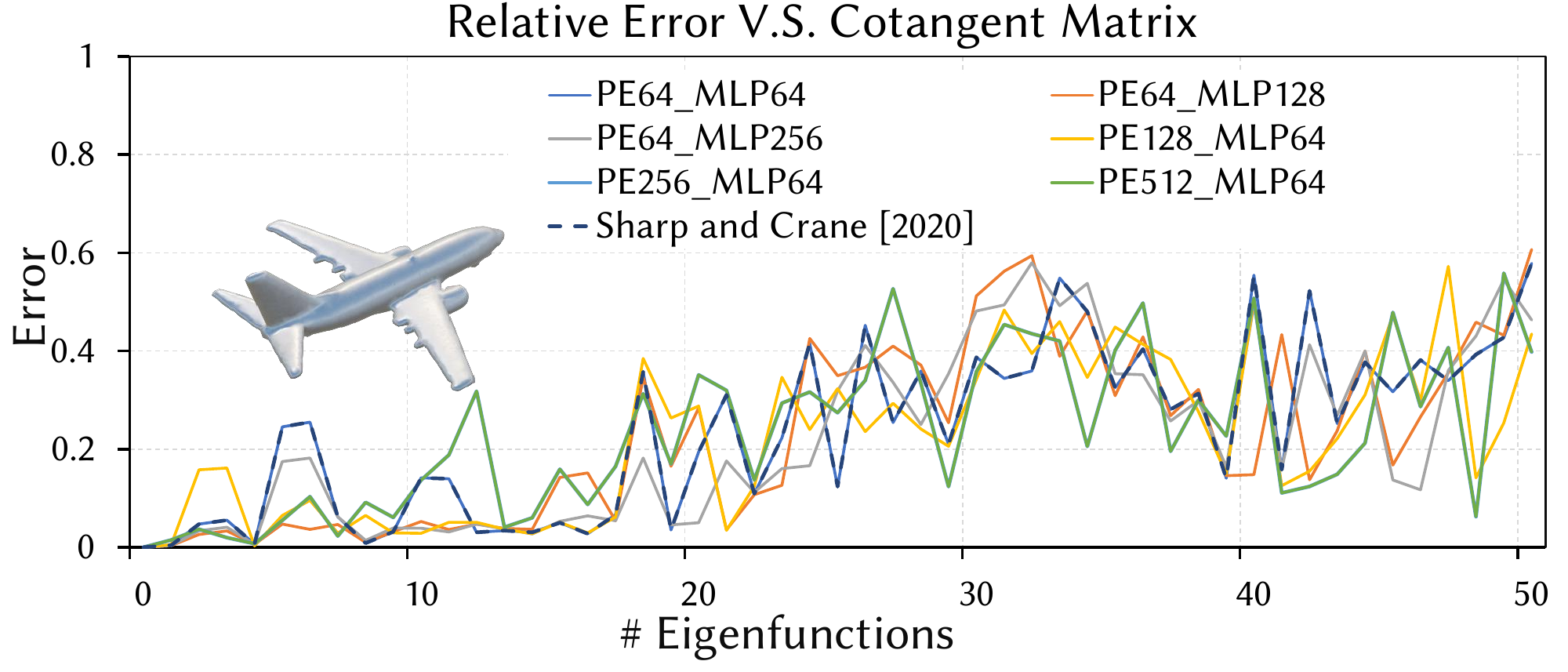}
\caption{Our method's error is comparable with the results of \citet{Sharp:2020:LNT}, who present a discretization-agnostic Laplacian eigenanalysis \Revision{on a single shape}. The different plotlines correspond to the execution of our method with different neural network configurations, with 'PE' indicating the maximum frequency of positional encoding and 'MLP' indicating the width of the MLP. }
\label{fig:error_number_eigenfunctions}
\end{figure}

\begin{figure}
\centering
\includegraphics[width=8cm]{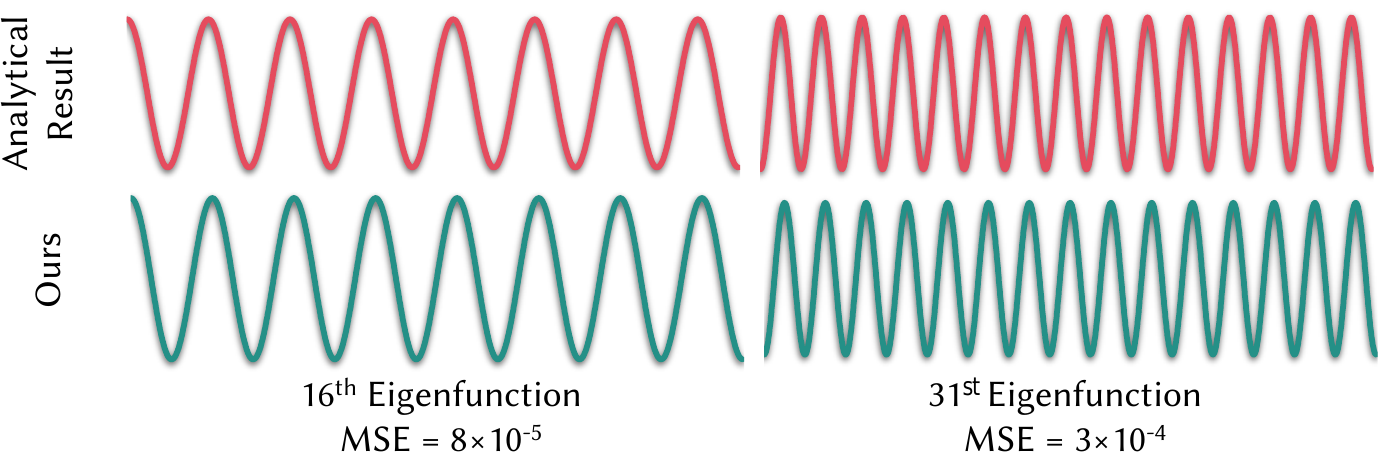}
\caption{For the 1D domain where the analytic solution is known, our method produces results that closely match the analytical solution. }
\label{fig:error_1D}
\end{figure}

\section{Experiments}

\subsection{Training Time \& Statistics}
We summerize the training time and network parameters for all examples in \reftab{timing-overall}, all data reported was obtained on on an AMD Ryzen 9 7950X CPU and an NVIDIA GeForce RTX 4090 GPU.

\begin{table*}
 \caption{Timing statistics. We report the training and inference times for all examples presented in this paper, along with the number of eigenfunctions and the parameters per MLP. The inference time represents the duration required for a single evaluation of all eigenfunctions at 30k cubature points. \Revision{We assume a uniform, fixed density for all examples.}}
  \centering
  \vspace*{-0.13in}

  \begin{tabular}{c|c|c|c|c}
    \toprule
    Example & Training Time & Inference Time(ms) & Eigenfunction Count & Parameters per MLP\\
    \midrule
    Bridge (\reffig{teaser})&  5.0h & 6 & 25 & 5.2k \\
    Teapot (\reffig{teapots})&  4.2h & 8 & 30 & 4.7k \\
    Airplane (\reffig{showErrorShape})  &  69s & 4 & 15 & 15k \\
    Round Chair (\reffig{new_chair_scalar}, \reffig{physicsPropertyTransfer})  &  4.6h & 3 & 10 & 15k \\
    Sofa (\reffig{new_chair_scalar})  &  4.0h & 3 & 10 & 15k \\
    Simulation on Wide Range of Shapes (\reffig{simulationGen})  &  35h & 5 & 15 & 13k \\
    Differentiable Sound (\reffig{sound})  &  6.9h & 11 & 32 & 4.6k \\
    Sound - Little Star (\reffig{SoundLittleStarMode}, \reffig{SoundLittleStar})  &  7.0h & 11 & 25 & 5.1k \\
    Walking robot (\reffig{LocomodeOpt1}, \reffig{LocomodeOpt2})  &  1.5h & 8 & 30 & 4.7k \\
    Walking Animals (\reffig{LocomodeOpt12animals})  &  19h & 7 & 30 & 5.0k \\
  \bottomrule
    
  \end{tabular}
  
  \label{tab:timing-overall}
\end{table*}

\subsection{Agreement with Discrete Operator Eigenfunctions}

In \reffig{error_number_eigenfunctions}, we evaluated the relative differences between our method and the cotangent matrix  as well as the analytical solution. For the first 20 eigenfunctions, our method's eigenfunctions exhibit an average difference of approximately $6\%$ compared to those from the cotangent matrix, increasing to $20\%$ as the number of eigenfunctions grows to 50. This error is comparable with \citet{Sharp:2020:LNT}, where both our method and theirs calculate Laplace eigenfunctions without the need for mesh connectivity. For the 1D domain, where the analytical solution is known, our method consistently produces results that closely match the analytical solution, as shown in \reffig{error_1D}.

\subsection{Rotation Invariance of Neural Eigenfunctions}

\begin{figure}
\centering
\includegraphics[width=8cm]{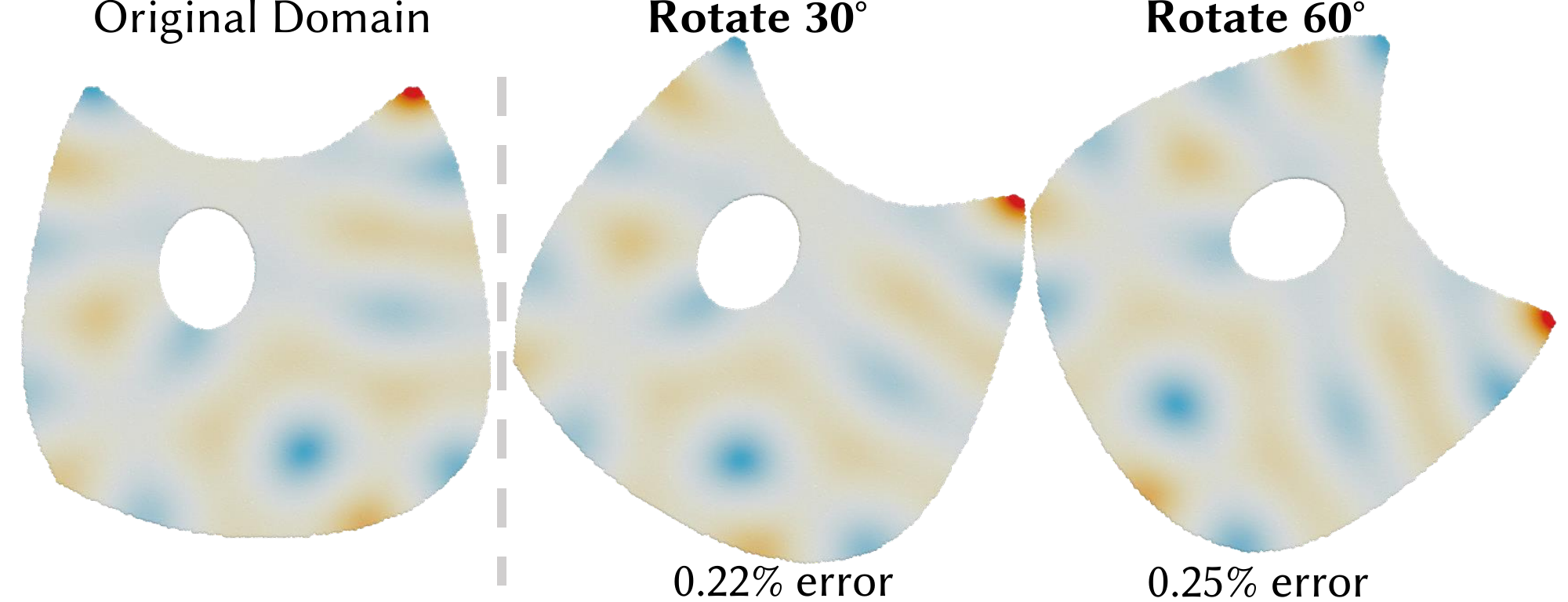}
\caption{
\textbf{Rotation invariance:} eigenmodes produced by our method are consistent with respect to rotation of the domain.} 
\label{fig:se3} 

\centering
\end{figure}

We retrained the network on rotations of the same shape with different angles. \reffig{se3} shows our method provides eigenfunctions under rotational perturbations that are highly consistent with the base unrotated configuration. Specifically, for rotations of 30 and 60 degrees, the average error over the first 30 modes between the rotated and original domains is $0.22\%$ and $0.25\%$, respectively. For context, variations within the $0.2\%-0.3\%$ range frequently occur due to re-initializing network weights before training, indicating no significant bias for rotations. The 30th eigenfunction from the three training results is visualized in this figure.

\subsection{Ablation Study on Causal Sorting}
\begin{figure}[h]
\centering
\includegraphics[width=\linewidth]{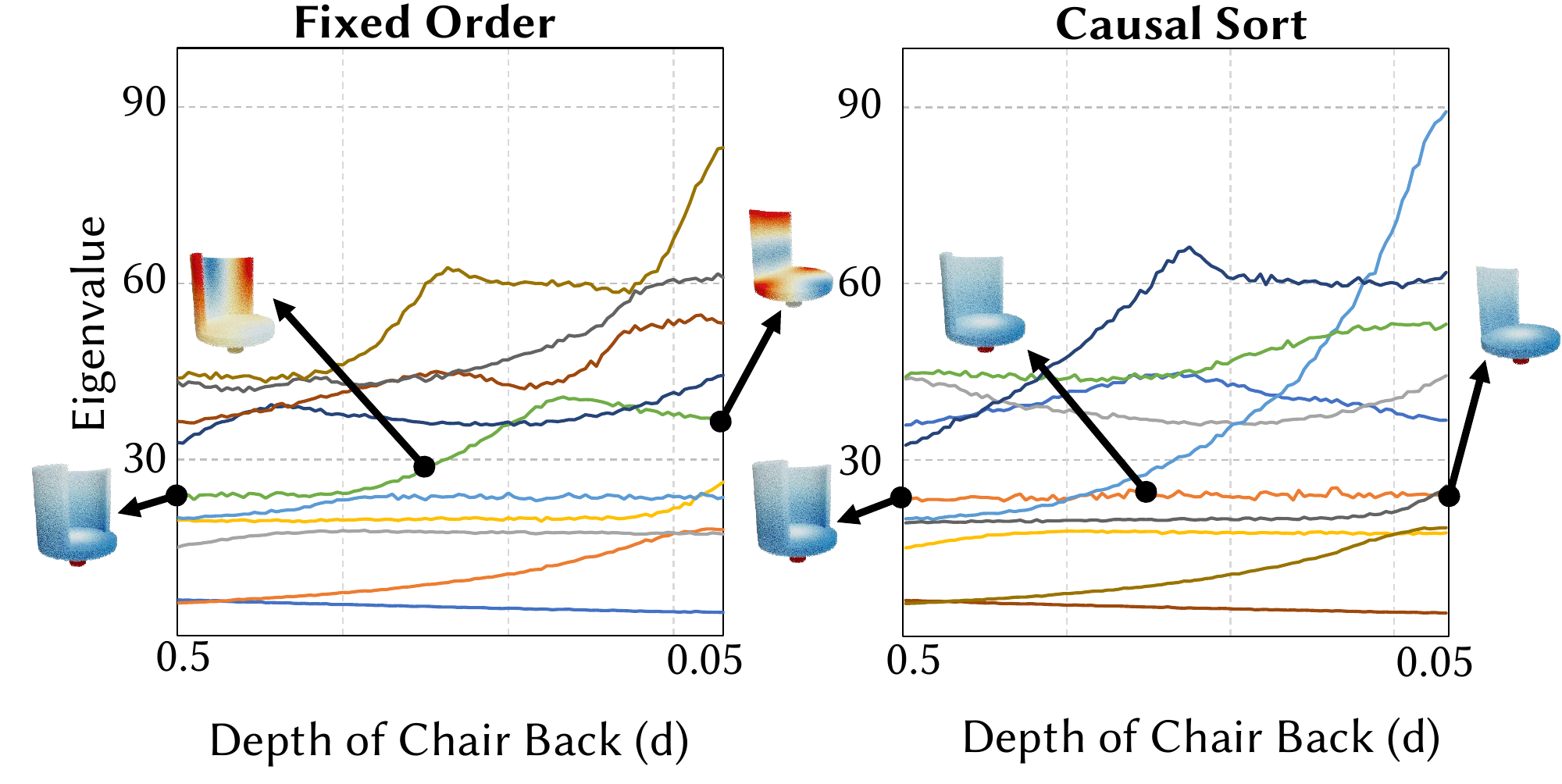}
\caption{
\emph{Eigenvalue Crossover.} Plot of eigenvalues versus geometry code for the chair shape space parameterized by radius ($r$), height($h$), and depth ($d$). comparing the na{\"\i}ve fixed-order method against shape-dependent causal sorting. Observe that both plots span a similar set of eigenvalues, but with different topology, with our proposed na{\"\i}ve baseline (\emph{left}) surpassed by our proposed method (\emph{right}), the latter more accurately resolving crossing eigenfunctions. Inherent to resolving an infinite function space with a finite number of eigenfunctions over shape space is that the highest-eigenvalue resolved eigenfunctions may include components of two or more unresolved eigenfunctions, in this case, as evident by the kink in the dark blue eigenfunction 7. For this plot, $r = 0.1$, $h = 0.8$, and $d$ varies.}
\label{fig:chair_crossover} 

\centering
\end{figure}

\begin{figure}
\centering
\hspace{-1cm}\includegraphics[width=8cm]{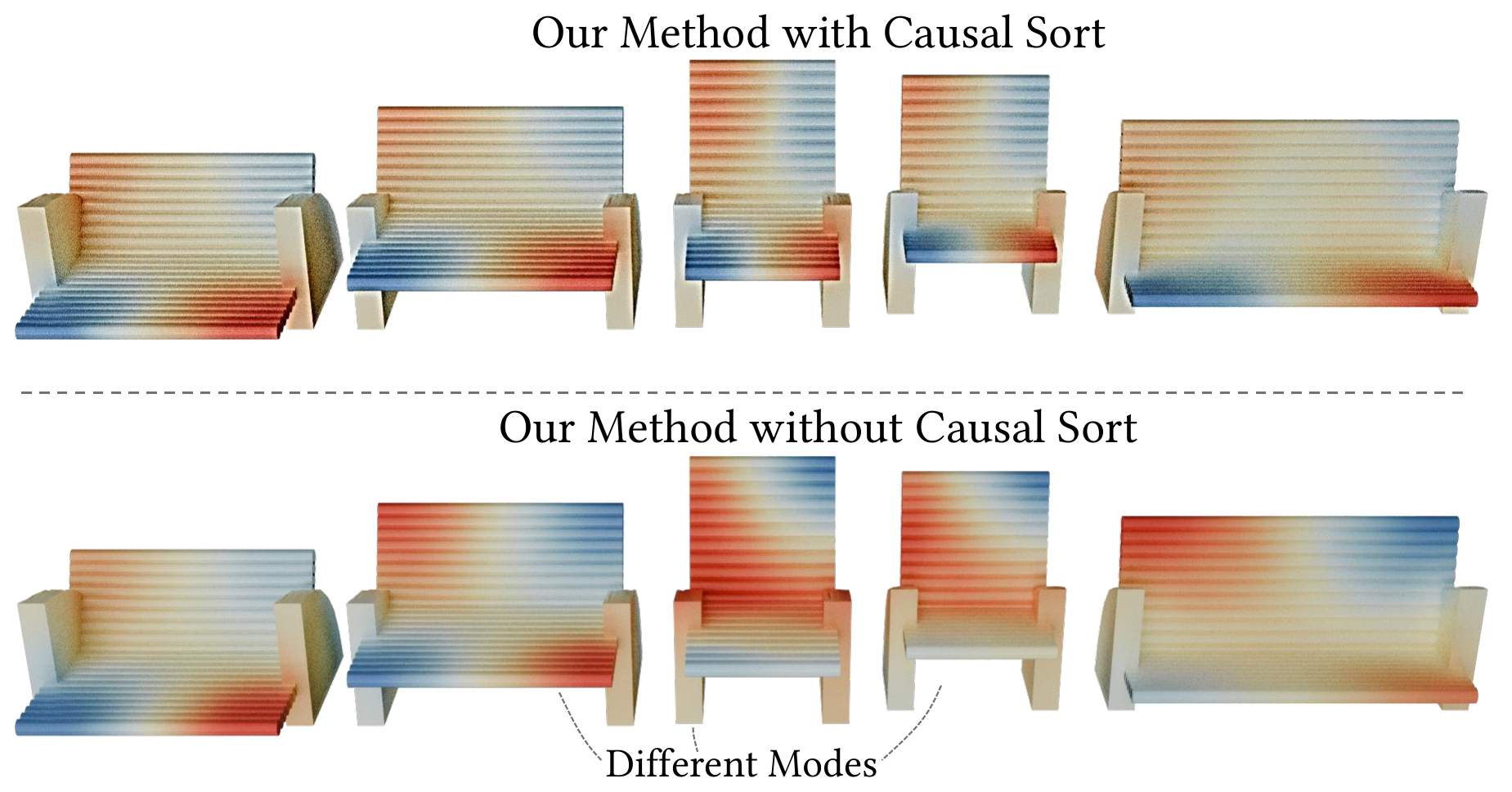}
\caption{
We visualized an eigenfunction from a simple chair shape space with and without causal sorting as we navigate the space. Our causal sorting approach preserve the mode pattern throughout shape space.
}
\label{fig:new_chair_scalar} 
\centering
\end{figure}

Figure \ref{fig:chair_crossover} plots eigenvalues with respect to geometry code for the shape space of Figure~\ref{fig:new_chair_scalar}.
Our method resolves eigenvalue crossings, whereas the proposed na{\"\i}ve baseline does not. The exception is the kink in the ``highest'' eigenvalue (eigenmode 7), which exposes a fundamental challenge in eigenanalysis over shape space (see \S\refsec{discussion}).


\subsection{Ablation Study on Consistent Modes}

The importance of resolving eigenvalue crossings is illuminated by studying the mode of an eigenfunction as it evolves over shape space. We've already in shown that resolving crossings leads to consistent mode patterns across shape space. In addition to that, as depicted in \reffig{new_chair_scalar} (top row), the absence of shape-dependent causal sorting leads to significant, rapid changes in the eigenfunction modes as the shape evolves over the kinks corresponding to unresolved eigenvalue crossings. By resolving crossings (bottom row), our method obtains a smooth evolution of eigenfunctions. 




\section{Results \& Applications}

\begin{figure}
\centering
\includegraphics[width=8cm]{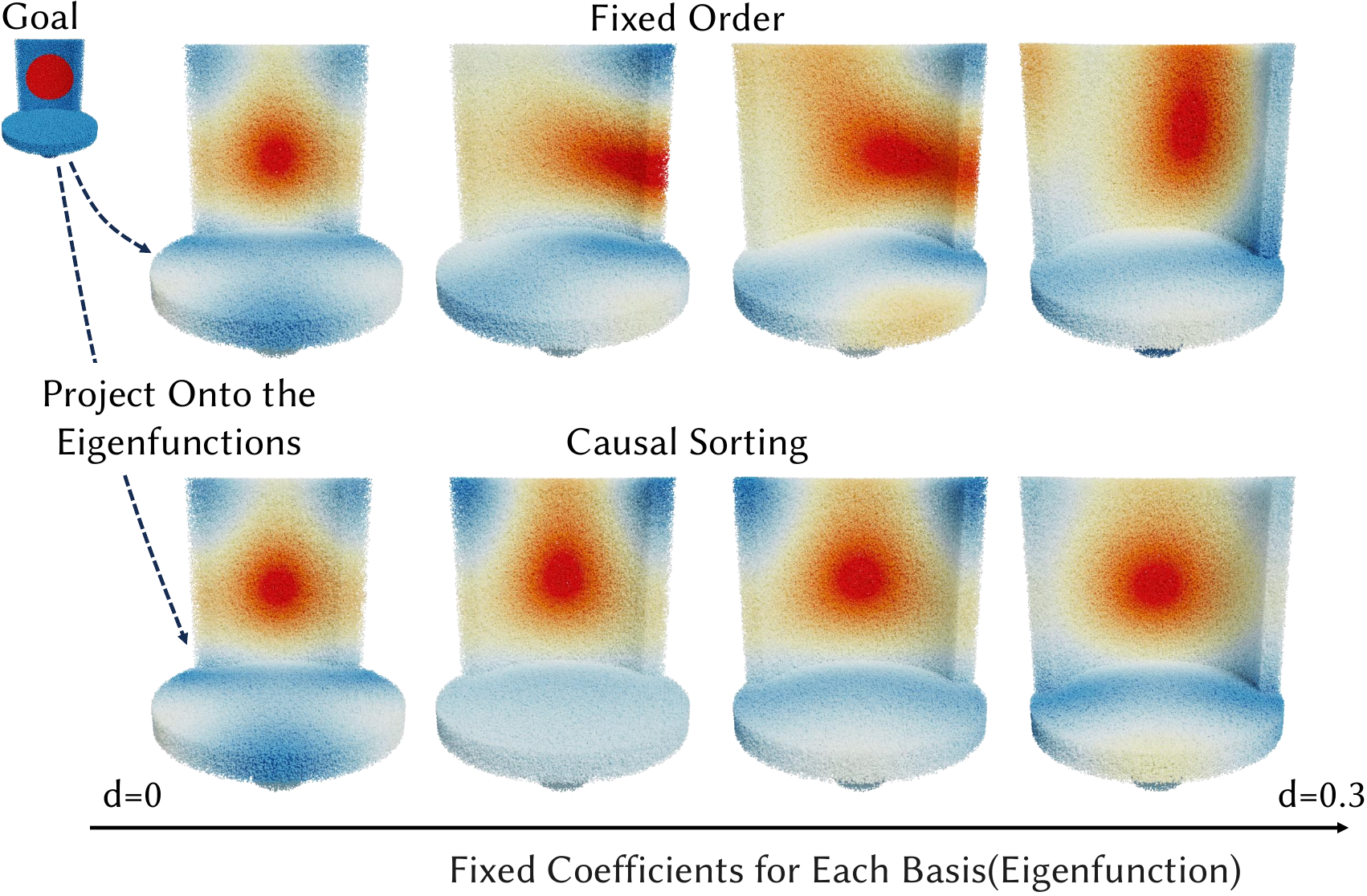}
\caption{
\emph{Physics Property Transfer.} A target heat distribution is projected onto the basis functions and transferred across the shape space. Causal sorting helps preserve initial distribution of the physical property.
} 
\label{fig:physicsPropertyTransfer} 
\centering
\end{figure}

\begin{figure}[h]
\centering
\includegraphics[width=8cm]{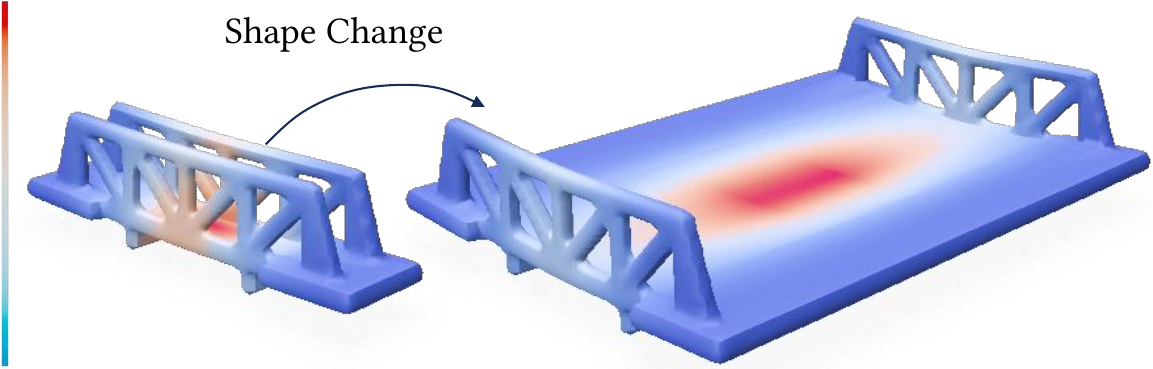}
\caption{
\emph{Interactive Visualization of Deformation.} The elastic modes of the bridge shape space enable interactive visualization of deformations across various shapes.
} 
\label{fig:simulationInteractiveElastic} 
\centering
\end{figure}

\begin{figure}[h]
\centering
\includegraphics[width=8cm]{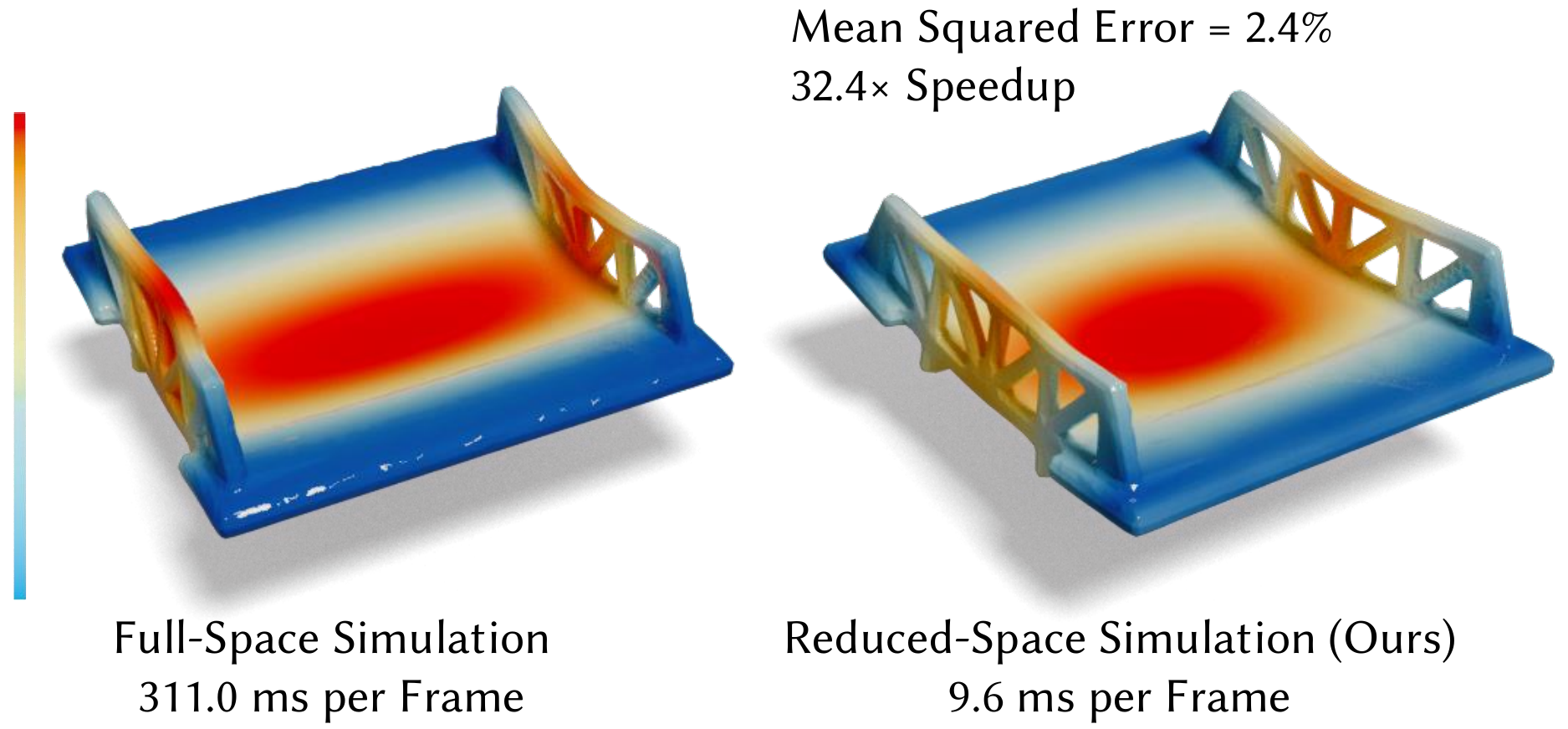}
\caption{
\emph{\Revision{Error and Runtime Comparison Between Our Method and Full-Space Simulation}} \Revision{We further evaluated the error in our reduced-space simulation by comparing it with a full-space simulation. Specifically, we simulated two bridges with fixed boundaries under gravitational loading, following the setup in \reffig{simulationInteractiveElastic}. The full-space simulation runs at 311.0 ms per frame. Our method achieves a runtime of 9.6 ms per frame, yielding a 32.4× speedup, with a mean squared error of 2.4\% in the final displacement. The norm of the displacement is shown using color coding.} 
} 
\label{fig:simErrorTime} 
\centering
\end{figure}

\begin{figure}[h]
\centering
\includegraphics[width=8cm]{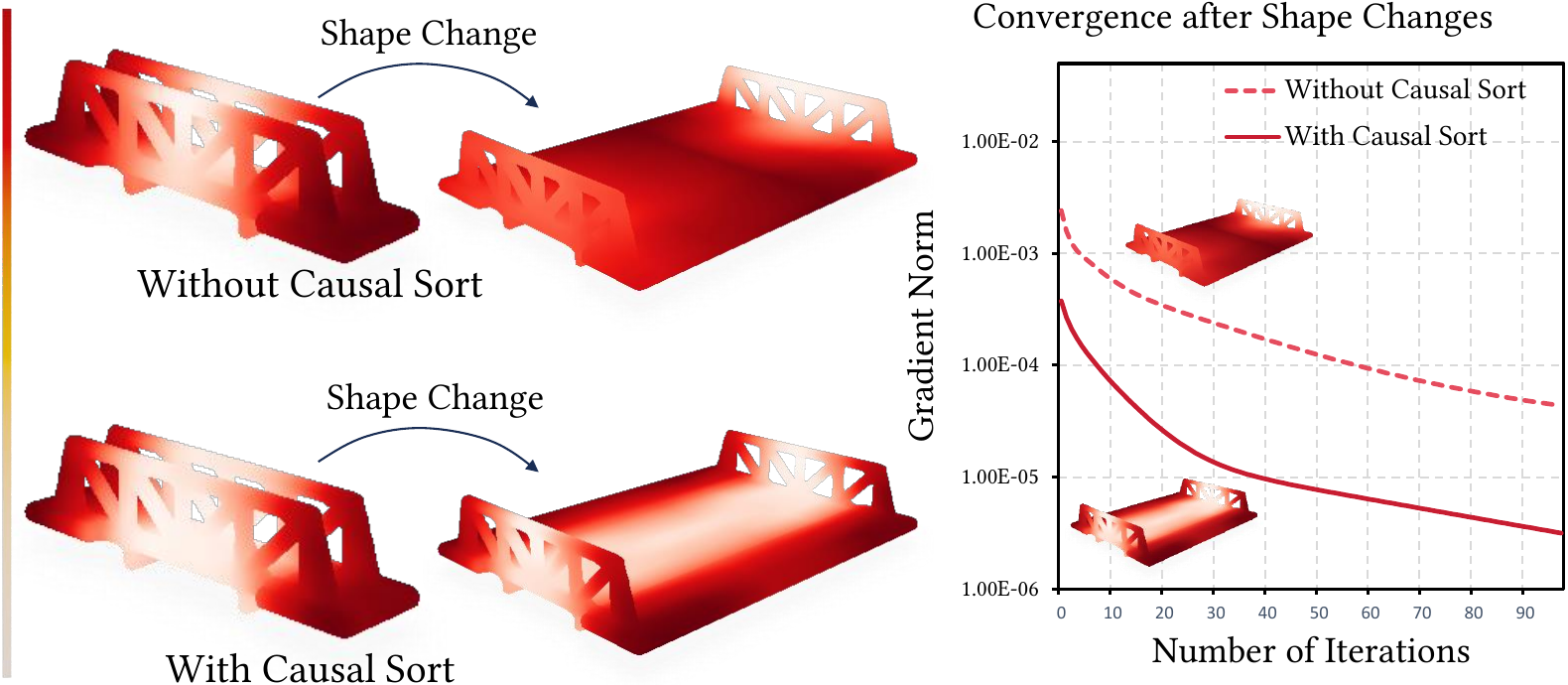}
\caption{
\emph{Interactive Visualization of Heat.} Once we solve the PDE for one shape, the solution can be transferred across the shape space, warm-starting the PDE solver over another shape. Without causal sorting, we observe an order of magnitude slower convergence, due to missed crossings. By ensuring consistent mode-shapes via causal sorting, our method enables faster convergence.
} 
\label{fig:simulationInteractiveHeat} 
\centering
\end{figure}

\begin{figure*}
\centering
\includegraphics[width=\textwidth]{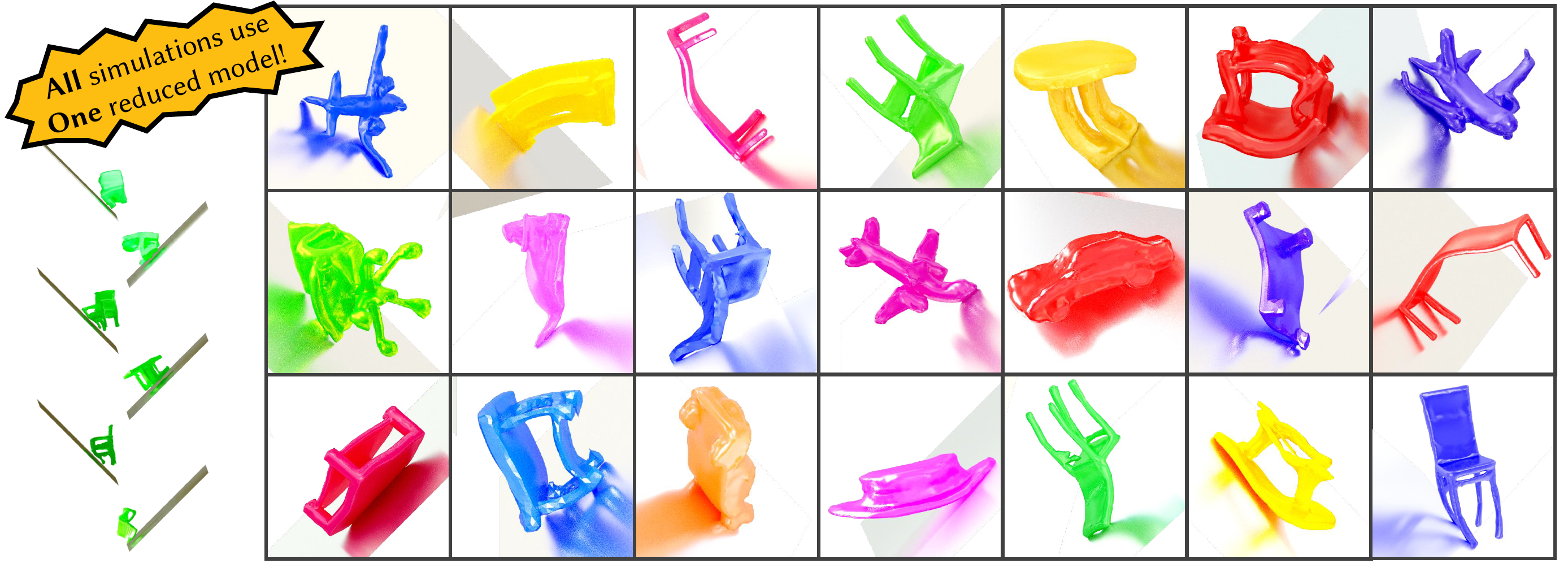}
\caption{
\emph{Reduced-space Simulation.} With just one training session, the model can represent the basis for hundreds of shapes, enabling direct reduced-space simulations. Using the pretrained occupancy field from \cite{chen2018implicit_decoder}, we constructed a shape space mapping a 256-dimensional coordinate to an occupancy field. After training on this shape space, our eigenfunctions can represent the skinning eigenmodes for all shapes within it. This allows for direct simulation of any shape from the model without requiring meshing over the spatial domain or retraining neural networks. 
} 
\label{fig:simulationGen} 
\centering
\end{figure*}

\subsection{Reduced-space Simulation for Shape Families} 
\paragraph*{Interactive Visualization of Physical Properties} 
Our shape-space eigenfunctions enable us to track how physical quantities evolve across the shape space, facilitating fast previews of physical properties. This significantly simplifies the often tedious and iterative shape design process. To demostrate this, we used the shape space shown in \reffig{teaser}, which is a 5-dimensional shape space that controls the width, length, height, fence size, and fence gap of a bridge. \Revision{We assume a uniform and fixed density for the bridge and all subsequent examples.} As shown in \reffig{simulationInteractiveElastic}, \reffig{simErrorTime} and \reffig{simulationInteractiveHeat}, the deformation and heat distribution initially defined on a simulated bridge can be seamlessly extrapolated to another bridge with a much larger width at runtime.

For instance, the solution $u(\bm{x})$ of a PDE computed over a shape $\bm{g}$ can be transferred to another shape $\bm{g}'$, providing a an excellent warm start $u'(\bm{x})$ for solving the PDE on the new shape: simply project onto the eigenbasis, $\alpha_i = \int_{\Omega_{\bm{g}}} u(\bm{x}) \phi^{\bm{g}}_i(\bm{x}) \, \mathrm{d}\bm{x}$, to obtain a low-dimensional embedding of the solution, then reconstruct on the new shape, $u'(\bm{x}) = \sum_i \alpha_i \phi^{\bm{g}'}_i(\bm{x})$. As illustrated in Figure~\ref{fig:simulationInteractiveHeat}, causal sort proves an order of magnitude lower error (and correspondingly faster convergence) using this warm start.

The reason for this is that the mode with causal sorting maintains greater consistency across the shape space, resulting in better preservation of the physical property distribution. As demonstrated in \reffig{physicsPropertyTransfer}, with causal sorting, the symmetric distribution remains intact as the shape changes. In contrast, without causal sorting, the distribution shifts unevenly across the chair's back.


 \begin{table*}[h]
    \begin{center}
        \caption{Quantitative comparisons on shapes from DiffSound \cite{Jin:2020:diffsound}. We evaluated the mean absolute error (MAE) by averaging the differences between predictions and ground truth across varying thicknesses. Our method enables pretraining on shape families, which is difficult for DiffSound due to its requirement for remeshing in each optimization step. This allows us to achieve a \Revision{compariable} MAE than DiffSound while significantly reducing query time.}
\begin{tabular}{l|cc|cc|cc|cc|cc|cccc}
\toprule

\multicolumn{1}{l}{Object} & \multicolumn{10}{c}{Target Thickness, \: {DiffSound\cite{Jin:2020:diffsound} /\colorbox[rgb]{0.9, 0.9, 0.9}{Ours}}}   & \multicolumn{2}{c}{MAE$\downarrow$}\\
\cline{2-11} 
\multicolumn{1}{l}{} &\multicolumn{2}{c}{0.3} & \multicolumn{2}{c}{0.4} & \multicolumn{2}{c}{0.5} & \multicolumn{2}{c}{0.6} & \multicolumn{2}{c}{0.7} &  \multicolumn{2}{c}{}\\

\midrule
Bunny &  0.304 &   \colorbox[rgb]{0.9, 0.9, 0.9}{0.299}      &     0.407  &  \colorbox[rgb]{0.9, 0.9, 0.9}{0.400}        &        0.508  &  \colorbox[rgb]{0.9, 0.9, 0.9}{0.497}      &      0.608  &  \colorbox[rgb]{0.9, 0.9, 0.9}{0.597}       &       0.709  &  \colorbox[rgb]{0.9, 0.9, 0.9}{0.699}   & {0.0073}   &  \colorbox[rgb]{0.9, 0.9, 0.9}{0.0016}\\
Armadillo &  0.338 &   \colorbox[rgb]{0.9, 0.9, 0.9}{0.284}      &     0.456  &  \colorbox[rgb]{0.9, 0.9, 0.9}{0.403}        &        0.590  &  \colorbox[rgb]{0.9, 0.9, 0.9}{0.501}      &      0.696  &  \colorbox[rgb]{0.9, 0.9, 0.9}{0.592}       &       0.730  &  \colorbox[rgb]{0.9, 0.9, 0.9}{0.691}   & {0.0623}   &  \colorbox[rgb]{0.9, 0.9, 0.9}{0.0075}\\
Bulbasaur &  0.308 &   \colorbox[rgb]{0.9, 0.9, 0.9}{0.298}      &     0.411  &  \colorbox[rgb]{0.9, 0.9, 0.9}{0.398}        &        0.512  &  \colorbox[rgb]{0.9, 0.9, 0.9}{0.500}      &      0.614  &  \colorbox[rgb]{0.9, 0.9, 0.9}{0.597}       &       0.718  &  \colorbox[rgb]{0.9, 0.9, 0.9}{0.695}   & {0.0125}   &  \colorbox[rgb]{0.9, 0.9, 0.9}{0.0023}\\
Squirtle &  0.312 &   \colorbox[rgb]{0.9, 0.9, 0.9}{0.296}      &     0.416  &  \colorbox[rgb]{0.9, 0.9, 0.9}{0.394}        &        0.520  &  \colorbox[rgb]{0.9, 0.9, 0.9}{0.495}      &      0.624  &  \colorbox[rgb]{0.9, 0.9, 0.9}{0.595}       &       0.718  &  \colorbox[rgb]{0.9, 0.9, 0.9}{0.707}   & {0.0177}   &  \colorbox[rgb]{0.9, 0.9, 0.9}{0.0054}\\
\bottomrule
\end{tabular}
    \label{tab:thickness_estimate_result}
    \end{center}
\end{table*}

\paragraph*{Direct Simulation on Generated Shapes} Recent methods represent shapes through mappings from a latent space to implicit representations, such as occupancy fields \cite{chen2018implicit_decoder} and signed distance fields \cite{Park:2019:DeepSDF}. By interpolating within this latent space, new shapes can be generated, and our method facilitates direct and efficient simulation on these generated models without requiring spatial domain meshing. We leveraged a pretrained model from \citet{chen2018implicit_decoder}, which provides a 256-dimensional shape space, training our approach on specific shape-space coordinates and testing it on additional generated shapes outside the training set.  

We implemented the method of~\citet{modi2024Simplicits}, which simulates corotational elasticity dynamics in a discretization-agnostic manner. They leverage the reduced basis proposed by~\citet{benchekroun2023FastComplemDynamics}, a linear blend skinning basis with skinning weights set to Laplace eigenfunctions. \Revision{In their formulation, the displacement in the reduced space simulation is expressed as a weighted sum of affine transformations applied to the shape’s rest positions, with the rigid mode captured by a single scalar weight for the entire shape.} These two works build a reduced model for a single shape. With our eigenanalysis, it is now possible to build a reduced model for a continuous family of shapes. 

As shown in Figure~\ref{fig:simulationGen}, we built one reduced model to simulate over 250 shapes from 13+ ShapeNet categories, including shapes outside the training set. With our single model, switching between these diverse shapes is as simple as selecting a new shape code, with no need for domain (re)meshing or retraining of neural networks. The resulting simulation effectively captures local deformations, as demonstrated by the deformed chair/desk legs due to collisions with the plane, shown in the right. 

To train this example, we first built the shape space, and then performed eigenanalysis. The shape space was trained as an implicit decoder mapping a 256-dimensional coordinate to an occupancy field~\cite{chen2018implicit_decoder}.  Eigenanalyzing this shape space yielded the Laplace eigenfunctions that serve as the scalar weight functions for the skinning eigenmodes~\cite{benchekroun2023FastComplemDynamics}. While this example uses a learned shape space, as we discussed earlier, our approach applies also to hand-designed shape spaces (e.g., parametric CAD models).

Traditionally, simulating a diverse set of objects requires meshing each individual generated shape and performing eigenmode calculations for discrete operators (matrices) on the resulting meshes. Depending on the meshing algorithm, creating a mesh can take anywhere from around 0.02 seconds \cite{gmsh} to over 20 seconds \cite{Hu:2018:TMW:3197517.3201353}, and the subsequent eigenanalysis further adds to the computational burden. For instance, at a resolution of 30k vertices, standard eigendecomposition using SciPy \cite{pauli2020scipy} takes over 6 seconds to compute the first 15 modes. In contrast, our method eliminates the need for meshing and directly computes eigenfunctions, enabling fast simulations across diverse shapes. It calculates eigenfunction for the same number of cubature points in just 0.005 seconds, significantly reducing both computational time and memory requirements.

Compared to prior discretization-agnostic methods such as \citet{modi2024Simplicits}, our pretraining time for this example is longer at 35 hours. However, our method generalizes to a broader range of shapes. While \citet{modi2024Simplicits} require retraining for each individual shape, our approach supports evaluation on a continuous family of shapes, making it more efficient when handling over 200 query shapes. Additionally, our method has a significant advantage in memory consumption, as its memory cost remains constant regardless of the number of shapes, whereas their method incurs a linear increase in memory cost as the number of shapes grows.

\subsection{Differentiable Modal Sound Synthesis}

 \begin{figure}
\centering
\includegraphics[width=8cm]{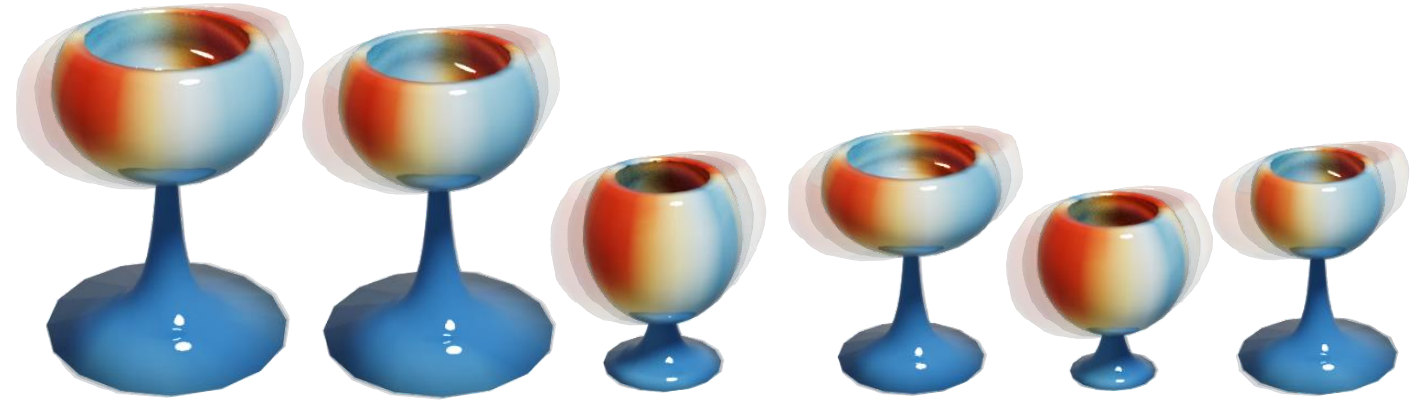}
\caption{
\emph{Elastic Mode Visualization.} We visualized the final shapes along with the elastic modes, using color coding to represent the norm of the $ \mathbb{R}^3$-valued eigenfunction. Note that the vibration is concentrated in the bowl of the glass, with almost no vibration occurring in the foot or stem.
} 
\label{fig:SoundLittleStarMode} 
\centering
\end{figure}

Our method enables fast shape optimization with eigenvalue-based objectives, such as finding a shape whose vibration modes and frequencies when struck produce a desired sound. We implemented the damped modal sound synthesizer and optimization approach proposed by \citet{Jin:2020:diffsound}. 
 \begin{figure}
\centering
\includegraphics[width=8cm]{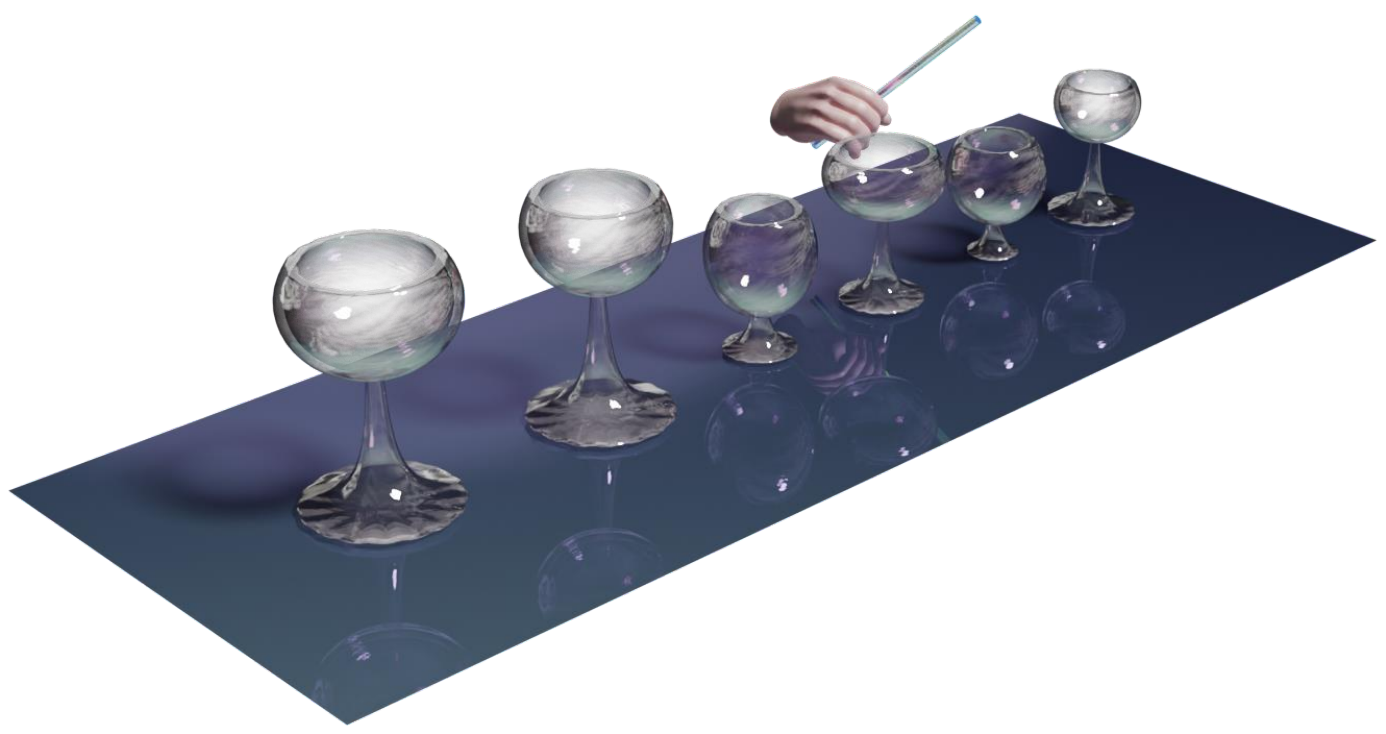}
\caption{
\emph{Little Star.} Our method enables the optimization of shapes to achieve target sounds. For instance, we optimize a family of wine glasses so that tapping their edges produces the melody of the song 'Little Star.'
} 
\label{fig:SoundLittleStar} 
\centering
\end{figure}

 \begin{figure*}
\centering
\includegraphics[width=\textwidth]{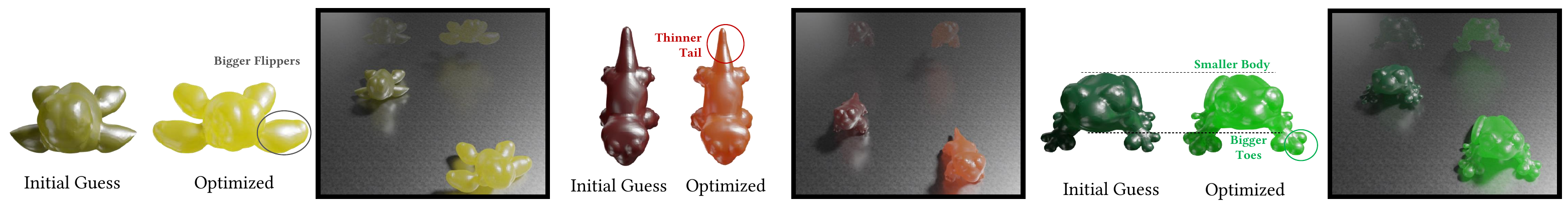}
\caption{
\emph{Shape Optimization for Locomotion Across a Broader Range of Characters.} We extended our approach to a more complex shape space derived from interpolations between 12 animals. This enhanced shape space enables the optimization of more complex and diverse shapes.
} 
\label{fig:LocomodeOpt12animals} 
\centering
\end{figure*}

They study the Volumetric Thickness Inference problem, which seeks to estimate the wall thickness $t$ of a hollow shape based on its sound when struck. 
We briefly recall their shape space definition: Given one solid shape, define a  family of hollow shapes parameterized by wall thickness $\bm{g} = t \in [0.15, 0.8]$, by thresholding the signed distance field (SDF) of the solid shape, such that $\bm{x}$ is in the shape if $-t < \text{SDF}(\bm{x}) < 0$. As a precomputation, we eigenanalyze the elastic energy Hessian for the shape space.

We implement their query: Given a target eigenvalue (vibration frequency) $\lambda_{i}^{\text{gt}}$, we seek the shape $\bm{g}^*$ that minimizes the difference in eigenvalue:
\begin{align}  \label{eq:sound}
\mathbf{g}^* = \underset{\mathbf{g}}{\arg\min} \, \sum_{i=1}^N \left\| \lambda_{i}(\mathbf{g}) - \lambda_{i}^{\text{gt}} \right\|^2 \ ,
\end{align}
where $\lambda(\bm{g})$ is the elastic energy, because the eigenfunctions are unit norm.

\begin{figure}
\centering
\includegraphics[width=8cm]{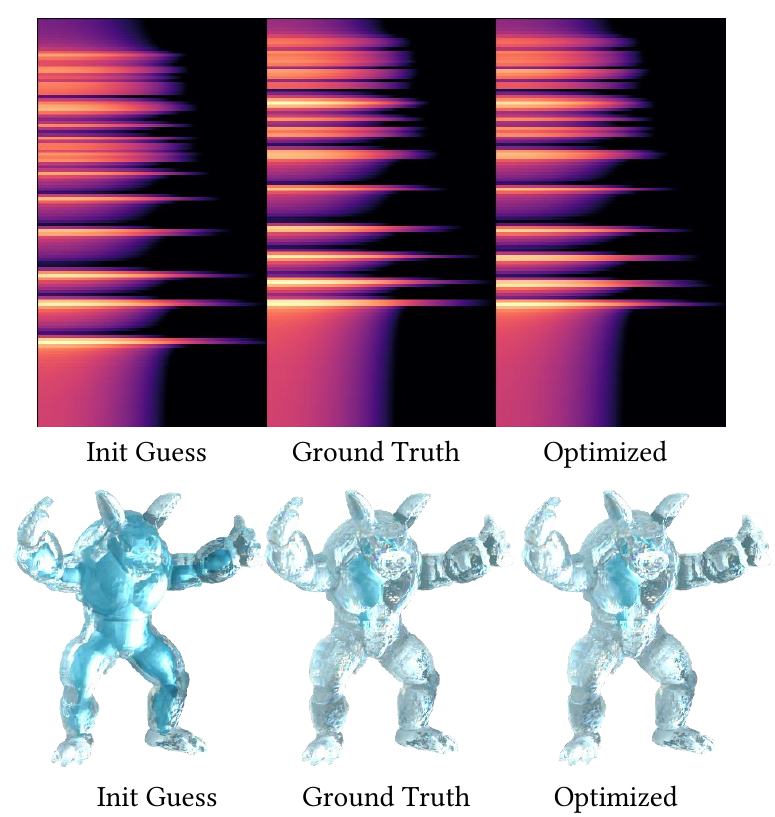}
\caption{
\emph{Volumetric Thickness Inference.} We optimize the thickness of the shape to produce sounds that better match the target sound. The spectrogram of the resulting sound and the corresponding shapes are visualized, showing that the optimized results are closer to the goal compared to the initial guess.
} 
\label{fig:sound} 
\centering
\end{figure}

 We follow their testing protocol: Select a target shape from four different shapes, select a thickness from $\{0.3, 0.4, 0.5, 0.6, 0.7\}$,  synthesize the desired (ground truth) sound based on the first $N=32$ eigenvalues, optimize \refeq{sound} and compare the thickness and spectrogram of the result $\mathbf{g}^*$. As depicted in \reffig{sound} the optimized shape and its spectrogram align more closely with the target shape and sound after optimization. As shown in \reftab{thickness_estimate_result}, our method achieves comparable error to \citet{Jin:2020:diffsound}, however, while they report about two hours per query, our approach completes in under three minutes, a $40\times$ speedup. Our discretization-agnostic approach avoids the costly remeshing required at each optimization step of their approach. Our method requires pretraining on the shape space, unlike \citet{Jin:2020:diffsound}. The precomputation time for our approach is approximately 7 hours per shape space, making it more efficient when handling 4 or more queries.

\paragraph{Shape Optimization for Pitch} We perform shape optimization with the goal of designing shapes that produce specific pitches. By optimizing the geometry of wine glasses and tapping them, we generate sounds corresponding to a predefined sequence of notes.

We constructed a 4-dimensional shape space for glasses, where the shape parameters control the radius of the bowl, the radius of the foot, the length of the stem, and a global scaling factor. We optimized the eigenvalue for a chosen low-frequency mode, which is visialized in \reffig{SoundLittleStarMode}. 

To ensure the chosen mode dominates the resulting sound, we set its damping factor, $d$, to 12.5, which is half the damping factor of the other modes; here $d$ 
is the damping factor in the model of Jin et al's Eqs. (9) and (10) \cite{Jin:2020:diffsound}.
We minimize the loss (see \refeq{sound}) over $N=1$ mode to obtain $\mathbf{g}^*$. 

To generate an engaging melody, we set the target eigenvalues $ \lambda^{\text{gt}} $ to correspond to the square of frequencies of the notes: 'do' ($ 261.63 \, \mathrm{Hz} $), 're' ($ 293.66 \, \mathrm{Hz}$), 'mi' ($ 329.63 \, \mathrm{Hz}$ ), 'fa' ($ 349.23 \, \mathrm{Hz}$ ), 'so' ($ 392.00 \, \mathrm{Hz}$ ), and 'la' ($ 440.00 \, \mathrm{Hz}$ ). Each optimization requires 500 gradient descent steps and takes less than 50 seconds to complete. As shown in Figure~\ref{fig:SoundLittleStar}, we successfully obtain shapes that achieve the target sound through optimization.

\subsection{Shape Optimization for Locomotion} 
Inspired by work showing the utility of eigenfunctions as an actuation signal, our method allows us to take locomotion optimization one step further, and ask questions regarding shape optimization for locomotion. Inspired by \cite{Benchekroun:2024:Actuators}, we define a controlled actuation force on a character their defined by their eigenfunctions. 
\begin{align*} \bm{d} = \bm{D}(\bm{g}) \bm{a}(t)\ , \end{align*}
where the matrix $\bm{D}(\bm{g}) \in \mathbb{R}^{3n \times 3r}$ contains the eigenfunctions of our elastic operator, specifically done by setting the columns $3i+1\ldots 3i+3$ to $[\phi_i^{\bm{g}}, \mathbf{0}_n, \mathbf{0}_n]^T$, $[\mathbf{0}_n, \phi_i^{\bm{g}}, \mathbf{0}_n]^T$, $[\mathbf{0}_n, \mathbf{0}_n, \phi_i^{\bm{g}}]^T$ respectively, for $0\leq i < r$ \cite{Tycowicz:2013:efficient_construcction}.



\Revision{
To move the character forward, we also added an explicit damping friction force $f_c \in \mathbb{R}^{3r}$ described by \citet{Benchekroun:2024:Actuators}, penalizing the relative velocity between the character and the ground,
 \begin{align*}
 f_c = \mu  J^c|_T \dot{z} \ ,
 \end{align*} 
where $\mu$ is the damping coefficient, and the contact Jacobian $J^c|_T$ maps the reduced velocity $\dot{z}$ to the full-space contact tangential velocities before projecting them back into the reduced space.
}




We couple this actuation signal with a differentiable simulation and run it for 200 timesteps and obtain the final full-space displacement field $\bm{u}^*(\bm{g}) = {\bm{D}(\bm{g})}{\bm{z}^*}$. 
Then, we optimize for the shape code parameters $\bm{g}^*$ by maximizing the norm of $\bm{u}^*$: \begin{align*} \bm{g}^* = \argmax_{\bm{g}} (\|\bm{u}^*(\bm{g})\|^2 ) \end{align*}

\begin{figure}
\centering
\includegraphics[width=8cm]{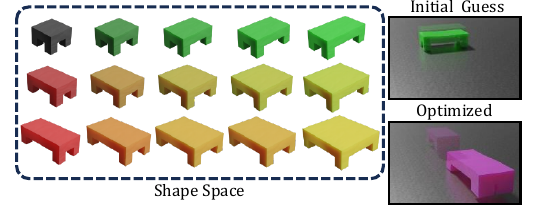}
\caption{
\emph{Shape Optimization for Walking Robot.} We optimize the design of a walking robot for a fixed controller, resulting in a robot that walks 18$\times$ faster than the initial design.
} 
\label{fig:LocomodeOpt1} 
\centering
\end{figure}

We first designed a 3-dimensional shape space for the walking robot. Starting from an initial guess within the shape space, we perform gradient-based optimization. The final shape obtained from our optimization moves faster than the initial guess, as demonstrated in Figure \ref{fig:LocomodeOpt1}. Note that this process is particularly challenging with traditional methods, as it requires computing the gradient of the eigenfunctions (which are used to design the controllers for the actuation forces) with respect to the shape parameters. This is difficult using traditional eigenanalysis based on matrices, and even more so for shapes with algebraic multiplicity.

\begin{figure}[h]
    \centering
    \begin{subfigure}[Visualization of gradient norm scaling in shape space. As shown on the left, fixed-order eigenfunction calculations result in mode changes at crossovers, leading to large gradient norms and making optimization more challenging. By contrast, the \Revision{middle} shows how our method, with causal sorting, achieves smoother eigenfunction transitions across the shape space, resulting in smaller gradient norms. \Revision{The consistent eigenfunction can lead to a better final solution of the optimization, as shown on the right.}]{
        \centering
        \begin{tikzpicture}[x=0.5\textwidth, y=0.5\textwidth]
            \node[anchor=south] (image) at (0,0) {
            \includegraphics[width=8cm]{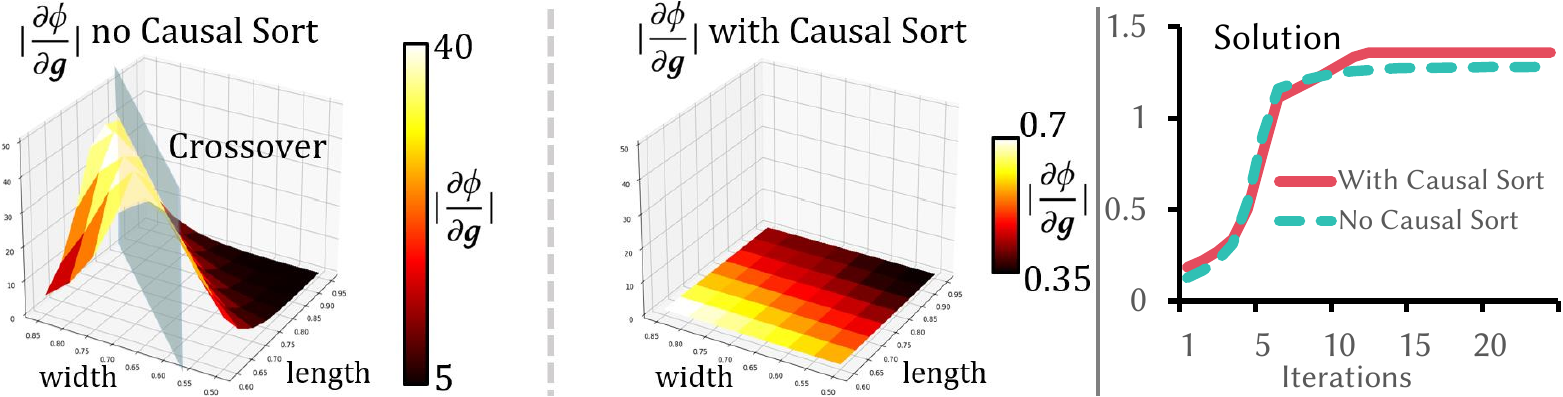}
            };
        \end{tikzpicture}
        \label{fig:sepsub1}}
    \end{subfigure}
    \hfill
    \begin{subfigure}[Starting from different initial guesses within the shape space of a walking robot, we observed two distinct final solutions with differing movement patterns when causal sorting was not applied. This occurs because the actuation force relies on eigenfunctions, and mode changes in eigenfunctions lead to variations in movement. By using causal sorting, we achieve consistent final solutions.]{
        \centering
        \begin{tikzpicture}[x=0.5\textwidth, y=0.5\textwidth]
            \node[anchor=south] (image) at (0,0) {
            \includegraphics[width=8cm]{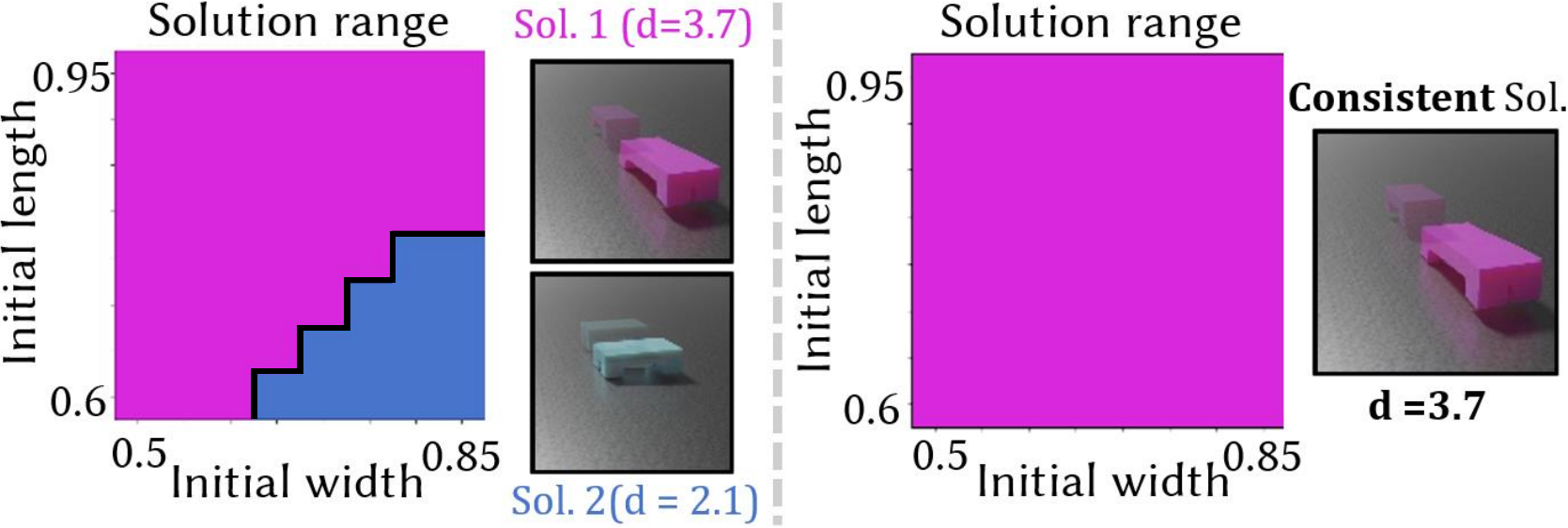}
            };
        \end{tikzpicture}
        \label{fig:sepsub2}}
    \end{subfigure}
    \caption{The mode changes at crossovers significantly increase the gradient norm in shape space, as shown in \ref{fig:sepsub1}. This creates an unsmooth gradient landscape, making optimization more challenging. Without causal sorting, as illustrated in \ref{fig:sepsub2}, the process resulted in two different final solutions. In contrast, causal sorting encourages consistent final solutions.}
    \label{fig:LocomodeOpt2}
\end{figure}

The change in modes across the shape space can impact the optimization process. Specifically, when modes shift, the gradient with respect to the shape space increases in magnitude, which can complicate the later stages of optimization. In Figure \ref{fig:sepsub1}, we visualize the shape-space norm gradient. Without causal sorting, these abrupt mode changes lead to a larger gradient, making the optimization process more challenging. In contrast, our proposed causal sort ensures that the modes transition smoothly across the shape space, as shown on the right, reducing this issue and enabling a more stable and efficient optimization process. 

\Revision{To demonstrate the impact of omitting causal sorting on optimization convergence, we performed shape optimization with and without causal sorting, using the same initial guess: width = 0.7, length = 0.7, and leg thickness = 0.3, optimized for 30 steps. As shown on the right of \reffig{sepsub1}, the final solution is better with causal sorting.} To further demonstrate this, we optimized the walking robot's shape with different initial guesses in the shape space and performed shape optimization, as shown in Figure \ref{fig:sepsub2}. Without causal sorting, two distinct final solutions are found, each with different movement patterns (in one, the robot walks sideways, and in the other, it walks vertically), with one solution corresponding to a local minimum. However, with causal sorting, only one consistent solution is obtained, with a stable walking pattern.

Our optimization approach can be extended to operate within a more intricate shape space derived from signed distance field (SDF)-based interpolation of 12 distinct animal shapes. This interpolation constructs a continuous 12-dimensional shape space, enabling smooth transitions between different animal characters and offering a rich variety of geometries for optimization. As illustrated in Figure \ref{fig:LocomodeOpt12animals}, our method effectively optimizes these animal characters within the shape space.

\section{Discussion and Future Work} \label{sec:discussion}

With these promising results, there remain areas for improvement and expansion. 

For a single shape represented by a triangle mesh, eigenanalysis of the Cotan matrix is  faster than stochastic gradient descent optimization. Our method's comparative advantage therefore lies in its generalization to shape spaces and general shape representations. 

As the number of eigenfunctions increases, so does their spatial frequency, and accurate training of higher frequencies is a known challenge for neural fields. Since our mathematical framework and optimization approach are not specific to neural fields, and it would be interesting to apply the variational formulation, causal gradient filtering, and causal sorting to the optimization of eigenfunctions in other representations, including representations that easily extend to higher frequencies.

\Revision{We have currently limited our scope to Neumann boundary conditions. Eigenfunctions subject to Dirichlet or Robin boundary conditions could be handled in the future by using PINN techniques for enforcing essential boundary conditions, such as blending with a weighting function that ensures constraint satisfaction by construction; alternatively, by introducing additional penalization terms that contribute to the loss function \cite{BERRONE2023e18820}. In the latter case, causal gradient filtering could be leveraged to ensure that the Dirichlet penalty dominates the variational energy corresponding to the PDE operator.}

Our implementation of stochastic cubature presently gives uniform weight to all samples. This is appropriate if the cubature is drawn from a uniform distribution. However, certain domain shapes may benefit from nonuniform cubature such as importance sampling. In that case, the number of samples per shape would potentially differ, potentially biasing the training to emphasize some regions of shape space over others, further necessitating a normalization of cubature weights per shape.

\begin{figure}
\centering
\includegraphics[width=1\linewidth]{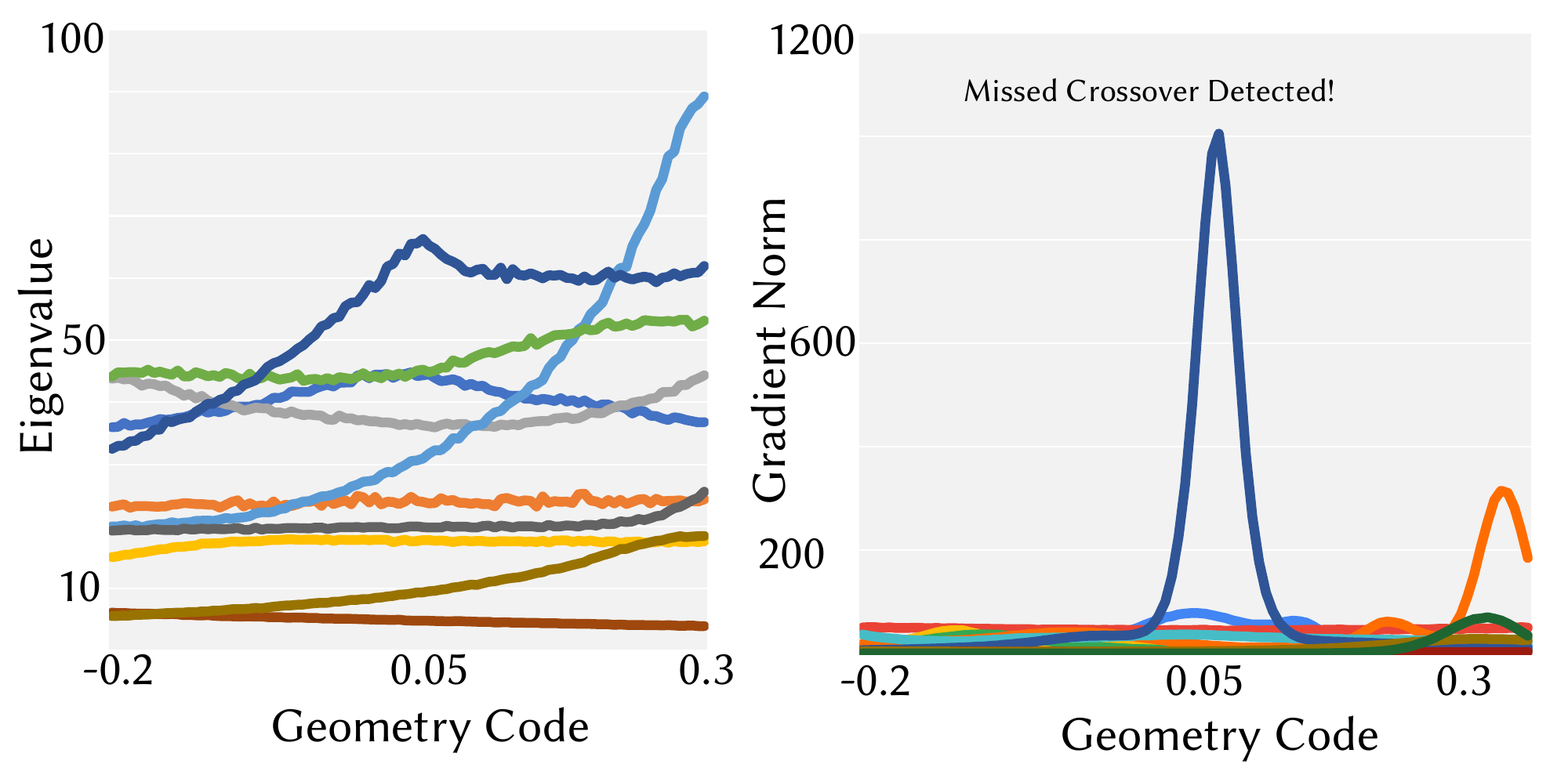}
\caption{
\emph{Gradient Indicators of Higher-Frequency Crossovers.} The norm of shape space gradients (the gradient with respect to the geometry code \(|\frac{\partial{\phi}(\bm{x}, \bm{g})}{\partial \bm{g}}|^2\) ) shows a significant increase at the crossover, therefore can be used as an indicator of the crossover from higher-frequency eigenfunctions. 
} 
\label{fig:shapecode_gradient} 
\centering
\end{figure}

Truncating an infinite-dimensional space with a finite number of subspaces inherently leads to missing information. In our case, this limitation can result in crossovers from higher-frequency eigenfunctions that cannot be fully captured due to the finite number of eigenfunctions used (recall eigenfunction 7 of Figure~\ref{fig:chair_crossover}). While with a finite budget we do not offer a fundamental way around this, we \emph{can} identify missed crossings by evaluating shape space gradients 
$\frac{\partial{\phi}(\bm{g}, \bm{x})}{\partial \bm{g}}$.
As shown in Figure~\ref{fig:shapecode_gradient}, there is a significant increase in the gradient norm at points where there is a crossover from a higher-frequency eigenfunction, which serves as a reliable indicator of these crossovers. In applications, then, the few eigenfunctions with missed crossings may be omitted, or, less aggressively, they may be included with knowledge of their piecewise nature. For example, in our application to transfer of PDE solutions across shapes, such an eigenfunction would be included if the two shapes lie on the same side of the crossing. 

In the future, it would be interesting to explore continuation methods~\cite{Allgower:Continuation} that seek to ``build out'' the remainder of the incomplete eigenfunction curves without arbitrarily increasing the total eigenfunction budget.

The gradients with respect to the geometry code may have further uses. For instance, by incorporating a shape space gradient norm into the training objective, it may be possible to achieve even smoother evolving functions among shape families. 

Shape-dependent representations of eigenfunctions appear to be a promising and versatile tool for working with eigenfunctions in continuously-parameterized shape spaces, broadening the scope of reduced physical models and accelerating inverse design problems. We believe that they also open the door to other applications in geometry processing (e.g., spectral processing methods), computer vision (e.g., shape correspondence), computational physics (e.g., fluid and deformable solid coupling), and indeed in all fields that leverage eigenanalysis of PDE operators.

\begin{acks}
We acknowledge the support of the Natural Sciences and Engineering Research Council of Canada (NSERC) grant RGPIN-2021-03733. We would also like to thank our lab system administrator, John Hancock, and our financial officer, Xuan Dam, for their invaluable administrative support in making this research possible. 
\end{acks}

\bibliographystyle{ACM-Reference-Format}
\bibliography{sample-bibliography}

\clearpage

\cleardoublepage

\end{document}


\title{Supplemental Document for Neural Representation of Shape-Dependent Laplacian Eigenfunctions}

\maketitle

\section{Proof of the Minimization of Dirichlet Energy Resulting in Vanishing Neumann Conditions}

The eigenfunctions $\phi(\bm{X})$ are derived by minimizing the Dirichlet energy under specific constraints. If these constraints are introduced via Lagrange multipliers, the eigenfunctions become stationary points of the functional $\mathcal{L}[\phi]$:

\begin{align*}
\mathcal{L}[\phi] =  \frac{1}{2}\int_{\Omega}\nabla \phi(\bm{X})^2 + 
 & \underbrace{\mu_1 \int_\Omega \phi(\bm{X})\phi_{prev}(\bm{X})}_{ \text{ Othorgnal constraint}}  +\\
\notag  & \underbrace{\mu_2 (\int_\Omega \phi(\bm{X})^2 - 1)}_{ \text{ Unit norm constraint}}
\end{align*}
where $\mu_1$ and $\mu_2$ are Lagrange multipliers, and $\phi_{prev}(\bm{X})$ represents the previous eigenfunctions. Since we have reached a stationary point, the following equation should be satisfied for \textbf{any} $\eta$:

\begin{align}
\mathcal{L}[\phi + \epsilon\eta] - \mathcal{L}[\phi] = \mathcal{O}(\epsilon^2) 
\label{eq:zero}
\end{align}

We can expand this equation as follows:
\begin{align*}
& \mathcal{L}[\phi + \epsilon\eta] - \mathcal{L}[\phi] = \\
& \underbrace{\frac{1}{2}\int_{\Omega} \left((\nabla\phi(\bm{X})+ \epsilon\eta)^2 - \nabla \phi(\bm{X})^2 \right)}_{\text{\ding{172}}} + \\
& \underbrace{\mu_1 \int_\Omega  \left((\phi(\bm{X})+ \epsilon\eta)(\phi_{prev}(\bm{X})+ \epsilon\eta) - \phi(\bm{X})\phi_{prev}(\bm{X}) \right)}_{\text{\ding{173}}} + \\
& \underbrace{\mu_2 \int_\Omega \left( (\phi(\bm{X})+ \epsilon\eta)^2 - \phi(\bm{X})^2\right)}_{\text{\ding{174}}}
\end{align*}

\begin{align*}
{\text{\ding{172}}} &= \frac{1}{2}\int_{\Omega} \left(\cancel{\nabla \phi(\bm{X})^2} + 2\epsilon\nabla \phi(\bm{X})\nabla \eta + \mathcal{O}(\epsilon^2) + \cancel{\nabla \phi(\bm{X})^2} \right) \\
&= \epsilon\int_{\Omega} \nabla \phi(\bm{X})\nabla \eta
\tikzmark{a}\\
&= -\epsilon\int_{\Omega} \nabla \phi(\bm{X}) \eta + \epsilon\int_{\partial\Omega} (\nabla \phi(\bm{X}) \cdot \bm{\mathsf{n}}) \eta \tikzmark{b}
\end{align*}
\begin{tikzpicture}[remember picture, overlay]
\draw[->] ([xshift=2mm] pic cs:a)
    to [out=0,in=0 ] 
     node[midway,anchor=west,xshift=3mm] {\footnotesize Integration by parts}
    ([xshift=2mm] pic cs:b);
\end{tikzpicture}

\begin{align*}
{\text{\ding{173}}}   &= \mu_1 \int_\Omega  \big(\cancel{\phi(\bm{X})\phi_{prev}(\bm{X})} + \epsilon\phi(\bm{X}) \eta + \epsilon\phi_{prev}(\bm{X}) \eta \\  
&  - \cancel{\phi(\bm{X})\phi_{prev}(\bm{X})} + \mathcal{O}(\epsilon^2) \big)\\
& = \epsilon\mu_1 \int_\Omega   \left(\phi(\bm{X})\eta + \phi_{prev}(\bm{X}) \eta\right)
\end{align*}

\begin{align*}
{\text{\ding{174}}}   &= \mu_2 \int_\Omega  \cancel{\phi(\bm{X})^2} +  2\epsilon\phi(\bm{X})\eta  + \mathcal{O}(\epsilon^2)  - \cancel{\phi(\bm{X})^2} \\
& = 2\epsilon\mu_2 \int_\Omega   \phi(\bm{X})^2 \eta
\end{align*}

We can rearrange the terms in $\mathcal{L}[\phi + \epsilon\eta] - \mathcal{L}[\phi] = {\text{\ding{172}}} + {\text{\ding{173}}} + {\text{\ding{174}}}$ and classify them into two types: volume integration $\mathbf{I}_{v}$ and surface integration $\mathbf{I}_{s}$.
\begin{align*}
\mathbf{I}_{v} = & 2\epsilon\mu_2 \int_\Omega   \phi(\bm{X})^2 \eta +  \epsilon\mu_1 \int_\Omega   \left(\phi(\bm{X})\eta + \phi_{prev}(\bm{X}) \eta\right) \\
& -\epsilon\int_{\Omega} \nabla \phi(\bm{X}) \eta
\end{align*}

\begin{align*}
\mathbf{I}_{s} = \epsilon\int_{\partial\Omega} (\nabla \phi(\bm{X}) \cdot \bm{\mathsf{n}}) \eta
\end{align*}

Since Equation \ref{eq:zero} holds for any $\eta$, we have:

\begin{align*}
    \mathcal{L}[\phi + \epsilon\eta] - \mathcal{L}[\phi] = \mathbf{I}_{v} + \mathbf{I}_{s} = 0
\end{align*}

We can construct $\eta$ such that it is zero at $\partial \Omega$ while being nonzero elsewhere. Therefore, the volume integration part $\mathbf{I}_{v}$ must be zero for any $\eta$. Given this conclusion, we can also infer that the surface integration part $\mathbf{I}_{s}$ holds for all $\eta$.

\begin{align*}
\mathbf{I}_{s} = \epsilon\int_{\partial\Omega} (\nabla \phi(\bm{X}) \cdot \bm{\mathsf{n}}) \eta = 0
\end{align*}

Finally, we have the vanishing Newmann conditions:

\begin{align*}
\int_{\partial\Omega} \nabla \phi(\bm{X}) \cdot \bm{\mathsf{n}}  = 0
\end{align*}

\section{Implementation Notes}

We nest the cubature scheme: each epoch considers one stochastically sampled  geometry code $\bm{g}_i \in \mathcal{G}$, and iterates over 
all position samples $\mathcal{X}_{\bm{g_i}}=\{\bm{x}_j \in \Omega_{\bm{g}_i}\}$:
\begin{align} 
\mathcal{L} = \underbrace{\sum_{\bm{g}_i\in \mathcal{G}}}_{\textrm{across epochs}} \ \ \underbrace{\sum_{\bm{x_j}\in \mathcal{X}_{\bm{g}_i}}}_{\textrm{within epoch}}
\nabla_{\bm{x}}\, \phi_i(\bm{g}_i, \bm{x_j})^2 \ .
\end{align}

\section{Ablation Study on Network Settings}
\begin{figure}
\centering
\includegraphics[width=8cm]{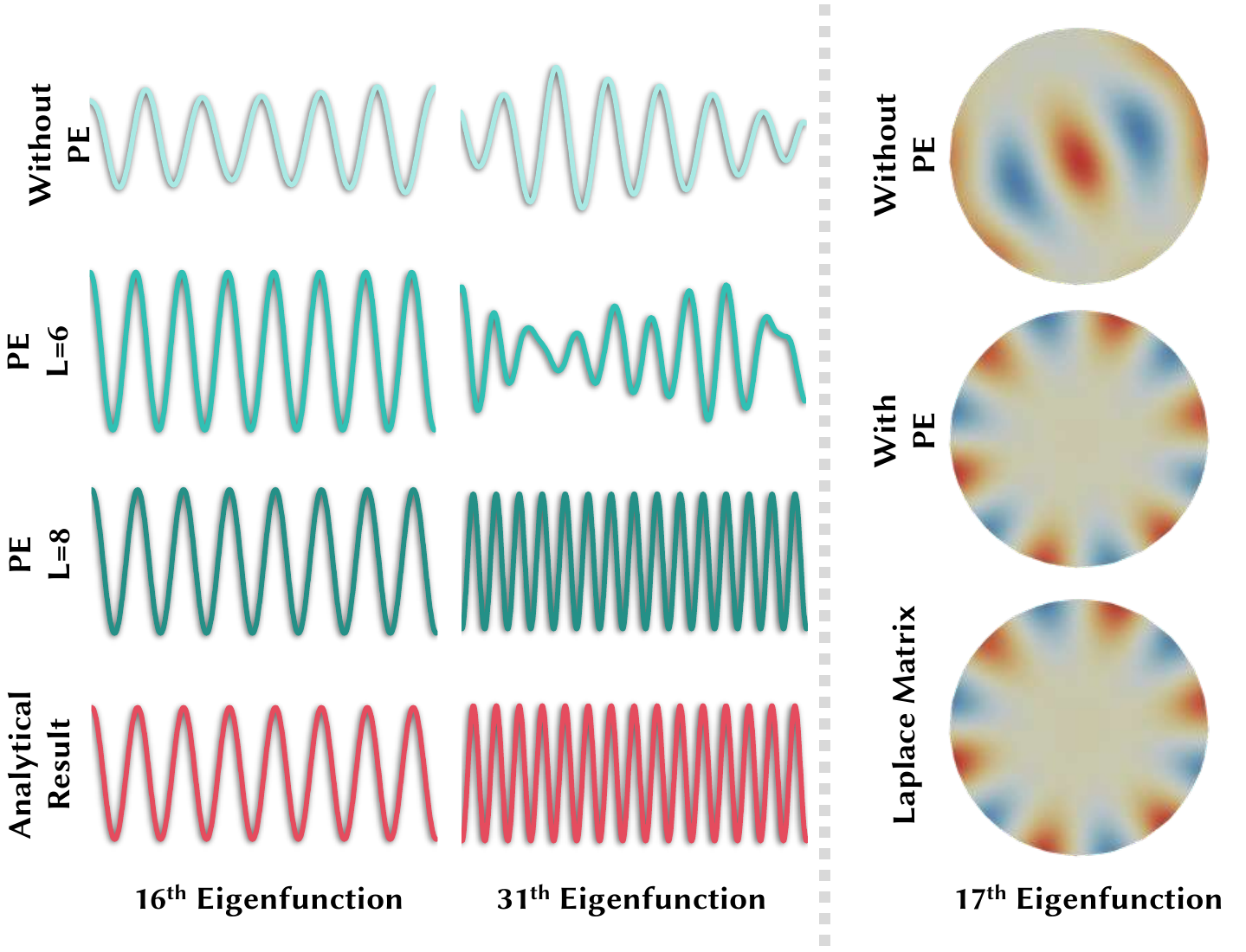}
\caption{
\emph{Positional Encoding.} We found that positional encoding (PE) aids the neural network in discovering high-frequency eigenfunctions. In the left part, we computed the eigenfunction of the Laplace operator in a 1D case, where we know the analytical result. Without PE or with a smaller $L$, the function struggled to capture the high-frequency eigenfunctions. Increasing $L$ (essentially increasing the maximum frequency in the encoded feature) can alleviate this issue.
In the right part, we calculated the eigenfunction of a 2D disk and compared the result with the eigenvectors from the Laplace matrix for sanity check. Without PE, the resulting pattern differs from the outcome of eigenanalysis on a Laplace matrix. However, with PE, we are able to find corresponding results.
\label{fig:positional_encoding} 
}
\centering
\end{figure}

\paragraph*{Positional Encoding.} Following prior works \cite{Mildenhall:2020:NeRF, Tancik:2020:Fourier}, we applied positional encoding(PE), and the encoded Fourier Features are used as the MLP's input. The basic idea is to map the input $\bm{X}\in\mathbb{R}^3$ to a higher-dimensional space $\gamma(\bm{x})\in \mathbb{R}^{6L}$ using a family of sine and cosine functions with increasing frequency:

\begin{align*}
\gamma(\bm{X})=\left(\sin(2^0\pi\bm{X}),\cos(2^0\pi\bm{X}),\dots,\sin(2^{L-1}\pi\bm{X}),\cos(2^{L-1}\pi\bm{X})\right)
\end{align*}
We observed that the parameter $L$ influences the learning of eigenfunctions with higher frequencies. In Figure \ref{fig:positional_encoding}, we analyzed the impact of $L$ on learning the 1D eigenfunction. The maximum frequency of the positional encoding (affected by $L$) affects the maximum number of eigenfunctions that our method can generate. With a larger $L$, our method finds it easier to reproduce high-frequency eigenfunctions. 

\paragraph*{Activation function.} Our loss function evaluates Dirichlet energy, which includes the first-order gradient term, necessitating the use of second-order gradients in the backward process. Consequently, the activation function for the MLP needs to be at least second-order differentiable (or $C^1$ continuous) to produce correct results. For instance, the ReLU activation function, despite its widespread use, is $C^0$ continuous and non-differentiable at zero. Additionally, the second-order gradient from autodiff for ReLU is always 0. Consequently, training an MLP with the ReLU activation function may fail to produce correct results for the second eigenfunction.

\begin{figure}
\centering
\includegraphics[width=8cm]{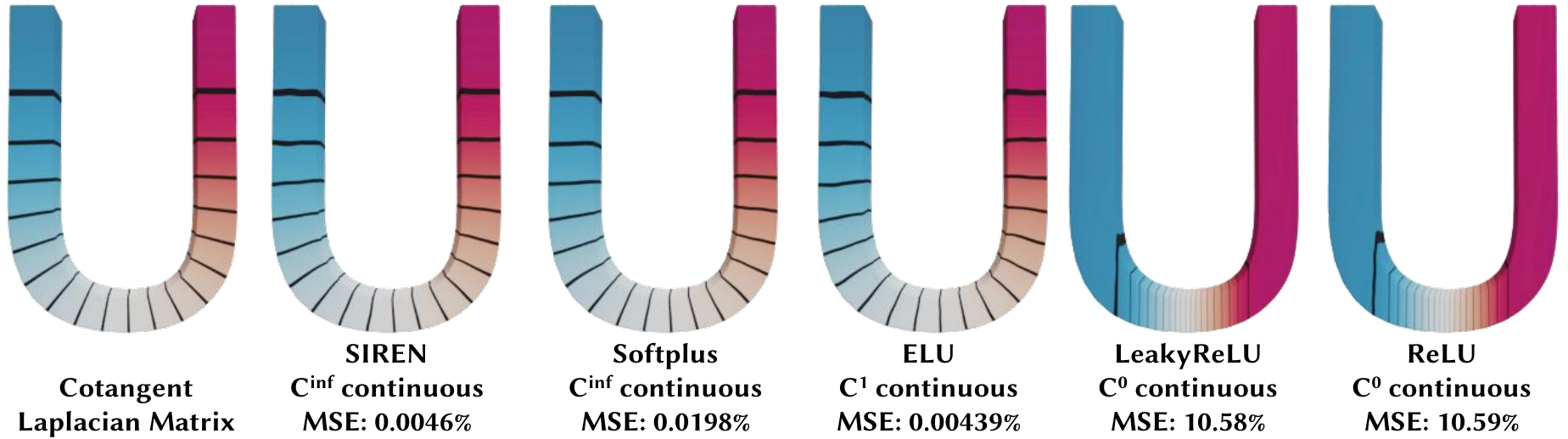}
\caption{
\emph{Activation functions.}
This image depict the second eigenfunction (or the Fiedler Vector in the discrete setting) of a "U" shape with different activation functions. The result from conducting eigenanalysis on a Cotangent Laplacian matrix is also included for sanity check. Since the loss function incorporates the first-order derivative, the gradient in the backward process computes the second-order derivative. Consequently, the activation function must be at least second-order differentiable (or $C^1$ continuous). All errors are evaluated against the second eigenvector of the Cotangent Laplacian Matrix. } \label{fig:Activation} 
\centering
\end{figure}

We conducted a comparison of resulting eigenfunctions using different activation functions with varying degrees of continuity in Figure \ref{fig:Activation}. We observed that activation functions with $C^1$ or higher continuity are capable of reproducing the second eigenfunction of a "U" shaped domain by penalizing Dirichlet energy, while others are not.

\section{Accuracy Evaluation in 1D case}

We've also evaluated the accuracy of our method in the 1D case, where the analytical result is already known. Specifically, the eigenfunctions for the Laplace operator in a 1D domain should be sinusoidal functions of increasing frequencies.

\begin{wrapfigure}{l}{0.25\textwidth}
    \centering
    \includegraphics[width=0.28\textwidth]{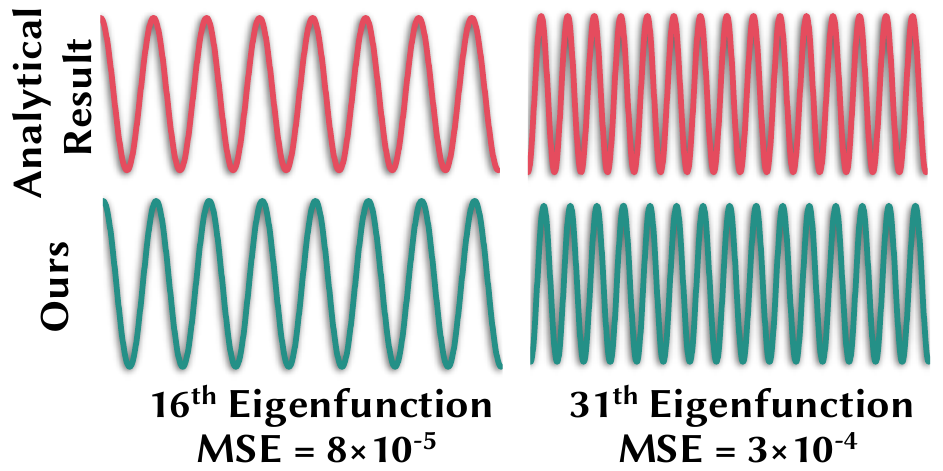}
\end{wrapfigure}

We computed the first 30 non-zero eigenfunctions by optimizing our network's weights. During each epoch, we randomly sampled 1500 cubature points within the domain to evaluate our gradient loss \ref{eq:loss}. The first eigenfunction, which is known to always be constant over the domain, was hard-coded to zero for simplifying the learning process. 

We evaluate the quality of our learned eigenfunctions by comparing them to the ground truth solution. %
To do so, we sample 1500 cubature points within our domain and measure their Mean Squared Error.
Because eigenfunctions may be unique up to a change in sign, we used the minimum MSE between the actual predicted function and its negative.
Our method successfully generates sinusoidal functions with the expected increasing frequencies matching the ground truth.
The MSE for all eigenfunctions was smaller than $5 \times 10^{-4}$, with an average MSE of $1.2 \times 10^{-4}$. 

\section{Additional Eigenvalue Plots for Shape-Dependent Causal Sorting}

\begin{figure*}
\centering
\includegraphics[width=\textwidth]{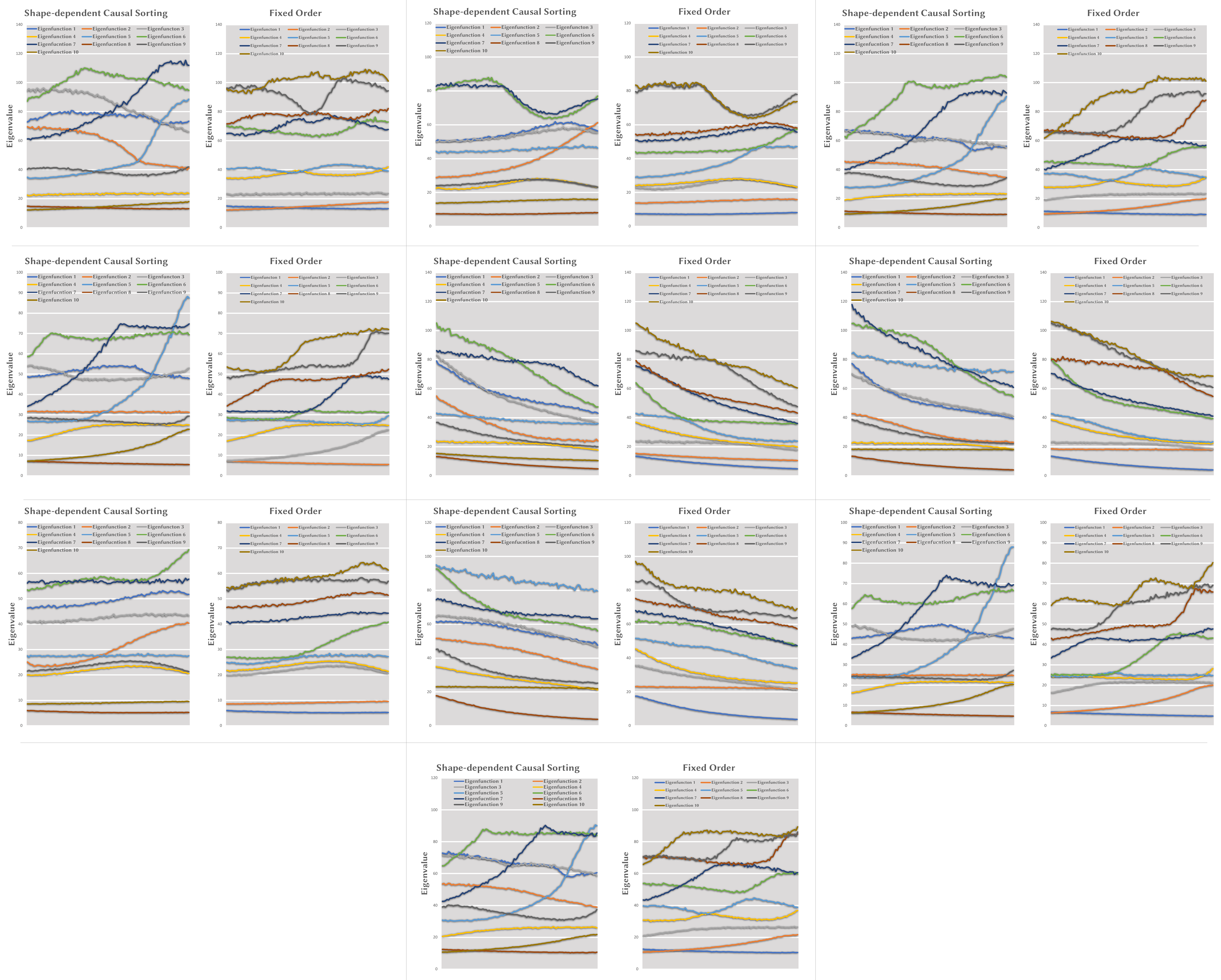}
\caption{
\emph{Another 10 Eigenvalue Plots.} To further demonstrate our method's ability to handle crossovers, we included 10 additional comparisons between our method and the naive fixed order approach. Using the eigenfunctions from the circle chair example, we sampled the shape domain 10 times and calculated the eigenvalues along one random direction. The left plots show our results, while the right plots depict the results from the naive method. In all 10 cases, our method enables crossovers that are not achievable with the naive approach.
 \label{fig:10Plots} 
}
\centering
\end{figure*}

To further demonstrate the efficiency of Shape-Dependent Causal Sorting, we included 10 additional eigenvalue plots. These plots show that our sorting method enables crossover in eigenvalues across different shape parameterizations. As shown in Figure \ref{fig:10Plots}, Shape-Dependent Causal Sorting achieves crossovers that are not possible with the naive fixed order approach in all these cases.

\bibliographystyle{ACM-Reference-Format}
\bibliography{sample-bibliography}